\documentclass{cta-author}
\usepackage{amsmath,amssymb,bm,mathrsfs,xfrac,mathtools,amsthm}
\newcommand*{\tran}{^{\mkern-1.5mu\mathsf{T}}}
\newcommand*{\hermconj}{^{\mathsf{H}}}
\usepackage{color}
\usepackage[hyphens]{url}
\usepackage[hidelinks]{hyperref}
\hypersetup{breaklinks=true}
\usepackage{mathptmx}
\usepackage{anyfontsize}
\usepackage{t1enc}
\usepackage{float}
\usepackage{tabularx}
\usepackage{multicol}
\usepackage[ruled,vlined]{algorithm2e}
\usepackage{siunitx}
\usepackage{soul}
\usepackage{array,graphicx}
\usepackage{booktabs}
\usepackage[table]{xcolor}
\usepackage{adjustbox}
\newcolumntype{R}[2]{>{\adjustbox{angle=#1,lap=\width-(#2)}\bgroup}l<{\egroup}}
\newcommand*\rot{\multicolumn{1}{R{90}{-.2em}}}
\usepackage[]{tcolorbox,xcolor}
\makeatletter
\newcommand*\bigcdot{\mathpalette\bigcdot@{.5}}
\newcommand*\bigcdot@[2]{\mathbin{\vcenter{\hbox{\scalebox{#2}{$\m@th#1\bullet$}}}}}
\makeatother
{}
{}
\newtheorem{remark}{Remark}{}
\newtheorem{definition}{Definition}{}
\definecolor{NREL}{RGB}{093,210,255} 
\newcommand{\norm}[1]{\left\lVert#1\right\rVert} 
\DeclareFontFamily{U}{mathb}{}
\DeclareFontShape{U}{mathb}{m}{n}{
<-5.5>mathb5
<5.5-6.5> mathb6
<6.5-7.5> mathb7
<7.5-8.5> mathb8
<8.5-9.5> mathb9
<9.5-11.5> mathb10
<11.5-> mathbb12}{}

\usepackage{ragged2e}
\usepackage{tikz}
\usetikzlibrary{positioning}
\usepackage{pgf,pgfplots}
\pgfplotsset{compat=1.15}
\usetikzlibrary{arrows}
\DeclareMathOperator*{\argmin}{arg\,min}
\makeatletter
\newcount\dirtree@lvl
\newcount\dirtree@plvl
\newcount\dirtree@clvl
\def\dirtree@growth{%
  \ifnum\tikznumberofcurrentchild=1\relax
  \global\advance\dirtree@plvl by 1
  \expandafter\xdef\csname dirtree@p@\the\dirtree@plvl\endcsname{\the\dirtree@lvl}
  \fi
  \global\advance\dirtree@lvl by 1\relax
  \dirtree@clvl=\dirtree@lvl
  \advance\dirtree@clvl by -\csname dirtree@p@\the\dirtree@plvl\endcsname
  \pgf@xa=0.3cm\relax
  \pgf@ya=-0.5cm\relax
  \pgf@ya=\dirtree@clvl\pgf@ya
  \pgftransformshift{\pgfqpoint{\the\pgf@xa}{\the\pgf@ya}}%
  \ifnum\tikznumberofcurrentchild=\tikznumberofchildren
  \global\advance\dirtree@plvl by -1
  \fi
}

\tikzset{
  dirtree/.style={
    growth function=\dirtree@growth,
    every node/.style={anchor=north},
    every child node/.style={anchor=west},
    edge from parent path={(\tikzparentnode\tikzparentanchor) |- (\tikzchildnode\tikzchildanchor)}
  }
}
\makeatother

\begin{document}
\title{\color{black}Measurement placement in electric power transmission and distribution grids: Review of concepts, methods, and research needs}
\author{\au{Marcos Netto}$^{\text{1}}$, \au{Venkat Krishnan}$^{\color{black}\,\text{2}}$, \au{Yingchen Zhang}$^{\text{1}}$, \au{Lamine Mili}$^{\,\text{3}}$}
\address{
\add{1}{Sensing and Predictive Analytics Group, Power Systems Engineering Center, National Renewable Energy Laboratory, Golden, CO 80401, USA}
\add{2}{\color{black}PA Consulting Group, Denver, CO 80203, USA}
\add{3}{\color{black}Bradley Department of Electrical and Computer Engineering, Virginia Polytechnic Institute and State University, Blacksburg, VA 24061, USA}
\email{marcos.netto@nrel.gov}}

\begin{abstract}
Sensing and measurement systems are quintessential to the safe and reliable operation of electric power grids. Their strategic placement is of ultimate importance because it is not economically viable to install measurement systems on every node and branch of a power grid, though they need to be monitored. An overwhelming number of strategies have been developed to meet oftentimes multiple conflicting objectives. The prime challenge in formulating the problem lies in developing a heuristic or an optimization model that, though mathematically tractable and constrained in cost, leads to trustworthy technical solutions. Further, large-scale, long-term deployments pose additional challenges because the boundary conditions change as technologies evolve. For instance, the advent of new technologies in sensing and measurement, as well as in communications and networking, might impact the cost and performance of available solutions and shift initially set conditions. Also, the placement strategies developed for transmission grids might not be suitable for distribution grids, and vice versa, because of unique characteristics; therefore, the strategies need to be flexible, to a certain extent, because no two power grids are alike. Despite the extensive literature on the present topic, the focus of published works tends to be on a specific subject, such as the optimal placement of measurements to ensure observability in transmission grids. There is a dearth of work providing a comprehensive picture for developing optimal placement strategies. Because of the ongoing efforts on the modernization of electric power grids, there is a need to consolidate the status quo while exposing its limitations to inform policymakers, industry stakeholders, and researchers on the research-and-development needs to push the boundaries for innovation. Accordingly, this paper first reviews the state of the art considering both transmission and distribution grids. Then, it consolidates the key factors to be considered in the problem formulation. Finally, it provides a set of perspectives on the measurement placement problem, and it concludes with future research directions.
\end{abstract}

\maketitle

\vspace{-.2cm}
\section*{Acronyms}

\vspace{-.2cm}
\noindent
\begin{tabular}{l l}
ARPA-E & Advanced Research Projects Agency-Energy \\
CT     & Current transformer \\
DER    & Distributed energy resource \\
DG     & Distributed generator \\
ILP    & Integer linear programming \\
IP     & Integer programming \\
LAV    & Least absolute value \\
PDC    & Phasor data concentrator \\
PMU    & Phasor measurement unit \\
PT     & Potential transformer \\
RTU    & Remote terminal unit \\
WLS    & Weighted least squares
\end{tabular}

\vspace{-.4cm}
\section*{Nomenclature}

\vspace{-.2cm}
\noindent
\begin{tabular}{l l}
$c$                 & Cost of installing a device on a given location \\
$c_{link,k}$        & Cost per unit of length of communications link $k$ \\
$d_{k}$             & Degree of bus $k$, i.e., number of branches incident to bus $k$ \\
$e$, $\bm{e}$       & Error scalar, error vector \\
$f_{k}$             & Frequency at bus $k$ \\
$\bm{g}(\bigcdot)$  & Vector-valued function of $\bigcdot$ \\
$i_{k\ell}$         & Complex-valued current phasor from bus $k$ to $\ell$ \\
$j$                 & $\sqrt{-1}$ \\
$m$                 & Number of measurements \\
$n$                 & Number of state variables \\
$n_{br}^{(k)}$      & Number of branches incident to bus $k$ \\
$n_{br}$            & Number of branches of an electric power network \\
$n_{bu}$            & Number of \emph{buses}, also called \emph{nodes}, of a power network \\
$n_{c}$             & Number of candidate locations for measurement placement \\
$n_{d}$             & Number of credible disturbances \\
\end{tabular}

\noindent
\begin{tabular}{l l}
$n_{link}$          & Number of communications links \\
$n_{p}$             & Number of placed measurements \\
$n_{pmu}^{(k)}$     & Number of PMUs that provide access to the voltage at bus $k$ \\
$p$                 & Probability index \\
$\bm{r}$            & Residual vector \\
$s$                 & Bus observability indicator equal to either $0$ or $1$ \\
$t$                 & Total time of a multistage placement \\
$u$                 & Measurement indicator equal to either $0$ or $1$ \\
$v_{k}$             & Complex-valued voltage phasor at bus $k$ \\
$\bm{w}$            & Vector of modal variables \\
$\bm{x}$            & Algebraic or dynamic state vector (see context) \\
$y_{k0}$            & Shunt admittance of bus $k$ \\
$y_{k\ell}$         & Series admittance between buses $k$ and $\ell$ \\
$\bm{z}$            & Measurement vector \\
$\bm{z}_{P}$        & Measurement vector containing real power flows/injections \\
$\bm{A}$            & Bus-to-bus connectivity matrix \\
$\bm{B}$            & Bus-to-branch connectivity matrix \\
$\bm{C}$            & Coherency matrix \\
$\bm{E}$            & Entropy matrix \\
$\bm{G}$            & Gain matrix \\
$\bm{H}$            & Measurement matrix \\
$\bm{H}_{P\theta}$  & $P\theta$ sub-matrix of the decoupled power flow Jacobian \\
$\bm{I},\bm{I}_{k}$ & Identity matrix, identity matrix of dimension $k$ \\
$J$                 & Objective function of choice \\
$N$                 & Number of \\
$\bm{R}$            & Error covariance matrix \\
$T$, $\bm{T}$       & Sampling period, transformation matrix \\
$\bm{U}$            & Matrix of right eigenvectors \\
$\bm{W}$            & Residual sensitivity matrix \\
$\bm{Y}$            & Admittance matrix \\
$\bm{Z}$            & Measurement matrix \\
\end{tabular}

\noindent
\begin{tabular}{l l}
$\mathbb{C}$        & Set of candidate locations for measurement placement \\
$\mathbb{D}$        & Set of credible disturbances \\
$\mathbb{B}$        & Set of all buses in an electric power grid \\
$\mathbb{E}$        & Set of all branches in an electric power grid \\
$\mathbb{I}_{0}$    & Set of all zero-injection buses in an electric power grid \\
$\mathbb{R}$        & Set of real numbers \\
$\mathbb{S}$        & Set of selected locations for measurement placement \\
$\mathbb{S}_{1}\cup\mathbb{S}_{2}$   & Union of the two sets $\mathbb{S}_{1}$ and $\mathbb{S}_{2}$ \\
$\mathbb{Z}$        & Set of integer numbers \\
$\mathcal{E}$       & Expectation operator \\
$\mathcal{I}$       & Information operator \\
$\mathcal{N}$       & Normally distributed random variable \\
$\alpha_{k\ell}$    & Phase angle of the complex-valued current phasor $i_{k\ell}$ \\
$\beta$             & Parameter of choice \\
$\gamma$            & System observability redundancy index \\
$\epsilon$          & Error bound \\
$\theta_{k}$        & Phase angle of the complex-valued voltage phasor $v_{k}$ \\
$\lambda_{k}$       & Length of communications link $k$ \\
$\mu$               & Sample mean \\
$\sigma_{k}$        & Standard deviation associated with measurement $k$ \\
$\bm{\upsilon}$     & Left singular vector \\
$\varnothing$       & Empty set \\
$\in$               & Element of \\
$\notin$            & Not element of \\
$\bm{1}_{k}$                         & Vector of all ones of dimension $k$ \\
$\otimes$                            & Kronecker product \\
\hspace{0.1cm}${\widehat{\bigcdot}}$ & Estimate of $\bigcdot$ \\
\hspace{0.1cm}${\tilde{\bigcdot}}$   & Measured $\bigcdot$ \\
$\bigcdot^{-1}$                      & Inverse of the matrix $\bigcdot$ \\
$\bigcdot\tran$                      & Transpose of the vector or matrix $\bigcdot$ \\
$\bigcdot\hermconj$                  & Conjugate transpose of the vector or matrix $\bigcdot$ \\
$|\bigcdot|$                         & Magnitude of a complex-valued variable $\bigcdot$ \\
$\text{card}(\bigcdot)$              & Cardinality, i.e., the number of elements, of a set $\bigcdot$ \\
$\text{det}(\bigcdot)$               & Determinant of a matrix $\bigcdot$ \\
$\text{Im}(\bigcdot)$                & Imaginary part of the complex-valued variable $\bigcdot$ \\
$\text{Re}(\bigcdot)$                & Real part of the complex-valued variable $\bigcdot$
\end{tabular}

\section{Introduction}

\vspace{-.2cm}
Sensing and measurement systems are ubiquitous in electric power grids. From power generation stations to end-customer sites, measurement systems are continuously acquiring raw data that are mission-critical to the long-term planning and real-time monitoring and control of power grids. In long-term planning, recorded measurements are used for model validation and calibration \cite{Hiskens2001} as well as for model reduction \cite{Chow2013}. Further, recorded measurements are used for the postmortem analysis of major events such as blackouts \cite{Taylor1997}. As for monitoring and control, sampled measurements are continuously feeding energy management systems running at control centers \cite{Wu2005} across the country. Sampled measurements enable near-real-time situational awareness, and corrective control actions are taken based on the analysis of the available information. Examples of the application for monitoring and control include static state estimation \cite{Abur2004}, security assessment \cite{Wehenkel1989, Krishnan2011}, event detection \cite{vonMeier2017}, and voltage control \cite{Kekatos2015}.

Despite their extension, experts recognize the need for further expanding the measurement systems in the near future \cite{GMLC2019}. One reason lies in the advent of new technologies---not only in sensing and measurement technology but also in communications and networking, optimization and control, energy conversion and storage, and power electronics---that are transforming the electric energy sector. The ongoing developments are being referred to as \emph{grid modernization}, and they are mainly driven by governmental initiatives aimed at the safe and reliable electrification of the economy. The Grid Modernization Laboratory Consortium,\footnotemark[1] supported by the U.S. Department of Energy, and the Global Power System Transformation Consortium,\footnotemark[7] are examples of such initiatives.
\footnotetext[1]{\url{https://gmlc.doe.gov/}}
\footnotetext[7]{\url{https://globalpst.org/}}

Grid modernization initiatives are pushing for rapid growth in the adoption of solar and wind power plants as primary sources of electric energy. Despite their advantages, however, these sources are naturally variable and stochastic \cite{Kroposki2017}. As a consequence, empirical data show that system states are experiencing excursions more often and more abruptly than in the past. Advanced controls are required to accommodate this variability and intermittency while maintaining grid reliability and enhancing grid resilience; in turn, additional measurements at higher spatial and temporal resolutions are required for advanced controls. This is arguably a worldwide trend.

In this scenario of the continuous expansion of measurement systems, the following question is posed:

\emph{Can the synergies across disparate measurement systems be exploited for an optimal placement solution that serves multiple applications?}

In pursuit of the answers, this paper reviews the prominent applications for which measurement placement strategies have been formulated. The paper also delves into the relatively newer applications for which measurement placement strategies have been formulated, particularly in the distribution grid because of the continuous increase in the number of distributed energy resources (DERs). The objective is to fulfill a gap in the open literature for a systematic framework guiding the placement of disparate measurement systems synergistically.

This study is motivated by the fact that the current practice in the industry hinges heavily on engineering judgment and the analysis of worst-case scenarios despite the extensive literature on the formulation of measurement placement strategies. For example, some of the most common industry practices include placing measurements to monitor critical devices/sites (e.g., step-up transformers in large power stations and tie-line buses) and to gather further information on specific locations to fix localized problems, such as voltage stability \cite{Madani2011}. On the other hand, academic publications tend to focus on specific applications. For example, there is much literature on the optimal placement of measurements to attain observability and thus enable static state estimation in transmission grids \cite{Manousakis2012}. Of course, these approaches are well justified within their own context. The industry needs a method that is reliable and cost-efficient yet simple, whereas academia mostly seeks novelty and rigor in the problem formulation. Accordingly, the primary role of this paper is to build a bridge between industry practices and existing formulations of the measurement placement problem. Further, this paper aims to review, present, and discuss measurement placement formulations in an accessible yet rigorous form. From a broader perspective, however, this paper serves multiple objectives, including to:

\vspace{-.3cm}
\begin{itemize}
\item Review, present, and discuss existing formulations for measurement placement in power grids. The review is thorough and rigorous though accessible to the larger community.
\item Identify synergies among different strategies that can lead to improved solutions. 
\item Inform the industry, policymakers, and researchers on the research-and-development needs for measurement placement.
\end{itemize}

\vspace{-.3cm}
Attaining these objectives is obviously challenging. First, the literature on this topic is vast; chronologically organizing key contributions on this topic is a laborious task. Second, the synergy among existing strategies is not obvious. To begin with, measurement placement is a classic planning problem, and it is challenging because the usage of measurements covers almost all the different types of power system research and analysis. Nonetheless, a multi-objective framework that offers a parsimonious solution to the contemporary problem of measurement placement for the future grid modernization scenarios and applications is required. The solution must recognize that future grids will include more uncertainty. Third, the identification of research-and-development needs that are worth the time and investment cost requires cross-disciplinary expertise.

{\color{black}
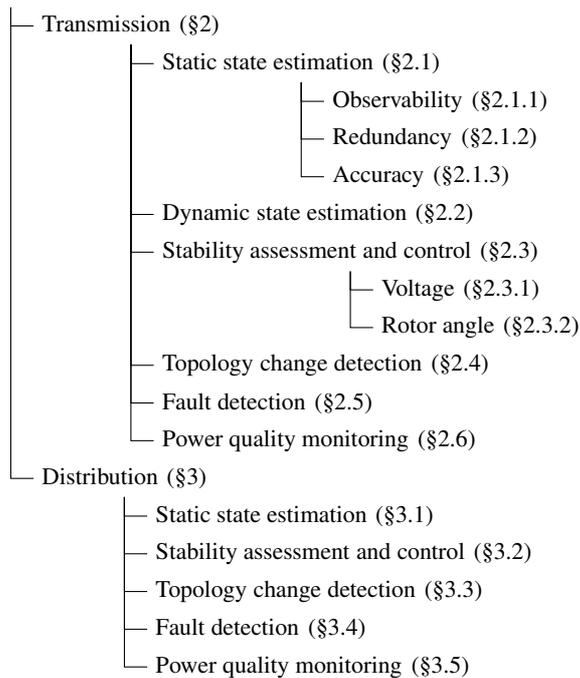
\begin{figure}[!t]
\centering
\begin{tikzpicture}[dirtree]
\node {Measurement placement} 
    child { node {Transmission $\,$(\S\ref{section.T})}
            child { node {Static state estimation $\,$(\S\ref{sec.StaticStateEstimationT})}
                child { node {Observability $\,$(\S\ref{subsec.ObservabilityT})} }
                child { node {Redundancy $\,$(\S\ref{subsec.RedundancyT})} }
                child { node {Accuracy $\,$(\S\ref{subsec.AccuracyT})} }
            }
            child { node {Dynamic state estimation $\,$(\S\ref{sec.DynamicStateEstimation})} 
            }
            child { node {Stability assessment and control $\,$(\S\ref{sec.StabilityAssessmentAndControlT})} 
                child { node {Voltage $\,$(\S\ref{subsec.VoltageT})} }
                child { node {Rotor angle $\,$(\S\ref{subsec.RotorAngle})} }
            }
            child { node {Topology change detection $\,$(\S\ref{sec.TopologyChangeDetectionT})} }
            child { node {Fault detection $\,$(\S\ref{sec.FaultDetectionT})} }
            child { node {Power quality monitoring $\,$(\S\ref{sec.PowerQualityMonitoringT})} }
    }
    child { node {Distribution $\,$(\S\ref{section.D})}
        child { node {Static state estimation $\,$(\S\ref{sec.StaticStateEstimationD})}
        }
        child { node {Stability assessment and control $\,$(\S\ref{sec.StabilityAssessmentAndControlD})}
            }
        child { node {Topology change detection $\,$(\S\ref{sec.TopologyChangeDetectionD})} }
        child { node {Fault detection $\,$(\S\ref{sec.FaultDetectionD})} }
        child { node {Power quality monitoring $\,$(\S\ref{sec.PowerQualityMonitoringD})} }
    };
\end{tikzpicture}
\caption{Categorization by application of the measurement placement problem in electric power grids.}
\label{fig.1}
\end{figure}
}

The reviewed measurement placement methods are separated into two sections---one for \emph{transmission grids} and the other for \emph{distribution grids}---to effectively address these challenges. See Fig. \ref{fig.1}. The methods are categorized within these sections by application (and sub-application, in some cases). For example, \S\ref{subsec.RedundancyT} presents and discusses methods aimed at enhancing the measurement redundancy for static state estimation in transmission grids. This categorization is critical in reviewing such extensive literature without losing sight of the main objectives previously outlined. {\color{black}Moreover, a categorization based on applications establishes common ground for comparisons between transmission and distribution grids.}

Note that the present work aims to provide insight into the works that made fundamentally new contributions to the literature or exposed an important factor to the problem of measurement placement; incremental contributions are referred to without further elaboration. Therefore, the \emph{References} section is not exhaustive---full cataloging of the existing literature on the present topic is beyond the scope of this paper. Second, before proceeding, we stress that many measurement systems acquiring nonelectrical quantities are found throughout electric grids. For example, equipment-level measurements of dissolved gas, vibration, temperature, pressure, humidity, and strain are used for equipment diagnostics. Measurements of solar irradiance and wind speed integrated with solar and wind power plants, respectively, are used in operational planning. Binary measurements of switchgear status are used for monitoring, topology processing, and control. In this paper, we focus primarily on measurement systems that acquire electrical quantities; within these, we are specifically interested in the measurements that support monitoring and control at a systems level. Table \ref{tab.1} summarizes the measurement systems of interest. Note that sample values acquired by merging units are popularly known in the United States as \emph{point-on-wave} measurements.

\begin{table}[t!]
\centering
\caption{Considered measurement systems and their location in the grid}
\begin{tabular}{@{}cr*{5}c}
\cmidrule[1pt]{1-7}
\cellcolor{white} & &
\rot{\parbox{60\unitlength}{Generation site\\1--25 kV}} &
\rot{\parbox{60\unitlength}{Transmission\\69--1,000 kV}} &
\rot{\parbox{60\unitlength}{Subtransmission\\15--69 kV}} &
\rot{\parbox{60\unitlength}{Distribution\\7.2--15 kV}} &
\rot{\parbox{60\unitlength}{Customer site\\100--240 V}}
\\
\cmidrule{2-7} 
\rowcolor{black!10} \cellcolor{white}
& Smart meter
$\;\;\,$&$\;\;\,$ 
$\;\;\,$&$\;\;\,$ 
$\;\;\,$&$\;\;\,$ 
$\;\;\,$&$\;\;\,$ 
$\;\;\,$&$\;\;\,$ \checkmark 
\\ 
& Power quality monitor 
$\;\;\,$&$\;\;\,$ 
$\;\;\,$&$\;\;\,$ 
$\;\;\,$&$\;\;\,$ \checkmark 
$\;\;\,$&$\;\;\,$ \checkmark 
$\;\;\,$&$\;\;\,$ 
\\ 
\rowcolor{black!10} \cellcolor{white}
& Digital fault recorder 
$\;\;\,$&$\;\;\,$ \checkmark 
$\;\;\,$&$\;\;\,$ \checkmark 
$\;\;\,$&$\;\;\,$ \checkmark 
$\;\;\,$&$\;\;\,$ \checkmark 
$\;\;\,$&$\;\;\,$ 
\\ 
& Digital protective relay
$\;\;\,$&$\;\;\,$ \checkmark 
$\;\;\,$&$\;\;\,$ \checkmark 
$\;\;\,$&$\;\;\,$ \checkmark 
$\;\;\,$&$\;\;\,$ \checkmark 
$\;\;\,$&$\;\;\,$ 
\\ 
\rowcolor{black!10} \cellcolor{white}
& PMU 
$\;\;\,$&$\;\;\,$ \checkmark 
$\;\;\,$&$\;\;\,$ \checkmark 
$\;\;\,$&$\;\;\,$ \checkmark 
$\;\;\,$&$\;\;\,$ 
$\;\;\,$&$\;\;\,$ 
\\ 
& Micro-PMU
$\;\;\,$&$\;\;\,$ 
$\;\;\,$&$\;\;\,$ 
$\;\;\,$&$\;\;\,$ 
$\;\;\,$&$\;\;\,$ \checkmark 
$\;\;\,$&$\;\;\,$ 
\\ 
\rowcolor{black!10} \cellcolor{white}
& Merging unit \cite{Song2017, Song2019}
$\;\;\,$&$\;\;\,$ \checkmark 
$\;\;\,$&$\;\;\,$ \checkmark 
$\;\;\,$&$\;\;\,$ 
$\;\;\,$&$\;\;\,$ 
$\;\;\,$&$\;\;\,$ 
\\ 
\cmidrule[1pt]{2-7}
\end{tabular}
\label{tab.1}
\end{table}

The most frequently adopted strategies to formulate the measurement placement problem include:

\vspace{-.2cm}
\begin{itemize}
\item Minimize the \emph{total cost}, including measurement devices and systems, infrastructure, data communications, storage, and processing.
\item Ensure \emph{system observability} under normal and anomalous operating conditions.
\item Maximize the \emph{performance of the applications} using the measurements, e.g., state estimation accuracy.
\end{itemize}

\vspace{-.2cm}
The first factor models the investment cost elements, whereas the other two capture technical aspects. The importance of including cost in the problem formulation is illustrated by an anecdotal example. Suppose that the operator of a power transmission grid of 900 buses decides to invest in making the measurement system observable solely by phasor measurement units (PMUs). The details of this problem, including the definition of observability, will be given in the next section. For now, it is sufficient to know that at least 300 PMUs are needed \cite{Baldwin1993}. Based on a report for the U.S. Department of Energy \cite{ARRA2014}, the cost of a PMU---including procurement, installation, and commissioning---can range from \$40,000--\$180,000, depending on the device class of precision, the number of measurement channels, and other characteristics; hence, the total investment in this anecdotal example ranges from \$12 million--\$54 million. Note that it is not uncommon to find transmission grids of this size around the world. The magnitude of the measurement placement in electric power grids---in terms of both problem dimension and involved costs---is formidable. 
{\color{black}
The following sections uncover additional technical factors. The contributions of this paper are as follows:

\vspace{-.3cm}
\begin{itemize}
\item 
It provides a comprehensive collection and an in-depth discussion of methods for measurement placement in electric power grids. 
\item 
It jointly discusses measurement placement in transmission and distribution grids. Note that even the most basic concepts, such as observability, have different interpretations for transmission and distribution grids, making this review a challenging undertaking. Such a review is nonexistent in the literature.
\item 
It reveals and discusses research gaps. For example, the discussion on leverage measurements is novel and not found elsewhere. The same applies to the discussion on dynamic state estimation, a timely and fast-evolving research area. The paper discusses several other research gaps in different application domains.
\item 
It identifies synergies among applications often considered separately. The idea of approaching the measurement placement problem from this standpoint is novel.
\end{itemize}
}

\vspace{-.3cm}
The paper proceeds as follows. Sections II and III present measurement placement strategies applied to electric power transmission and distribution grids, respectively. Section IV provides an outlook on what has been accomplished and delineates the major future research directions in this field. Section V concludes the paper.

\vspace{-.3cm}
\section{Measurement placement in transmission grids}\label{section.T}

\vspace{-.3cm}
In legacy transmission grids, all major substations are equipped with remote terminal units (RTUs); their role is to gather the measurements collected at the substation and transmit them to centralized control centers. Typically, RTUs transmit a batch of measurements once every 1--4 seconds. This paper does not consider the placement of RTUs and associated sensing and measurement devices. This is in part because another device, based on more recent technology and referred to as a PMU, has taken precedence in transmission grids. Depending on their class and manufacturer, PMUs can transmit a batch of Global Positioning System-synchronized measurements up to 240 times per second \cite{vizimax}. On a historical note, PMUs became commercially available a few years after the development of the prototype was completed in 1988 \cite{Phadke1993, Phadke2002}. The U.S. Department of Energy released the wide-area measurement system project \cite{Mittelstadt1995, Mittelstadt1996} shortly after that. The placement of PMUs has been well explored since then, with initial work dating to the early 1990s. A review of PMU placement methodologies developed up to 2011, with a focus on static state estimation only, is given in \cite{Manousakis2012}. In a complementary fashion, this paper encompasses all the different applications that can be considered for PMU placement, starting with static state estimation.

\vspace{-.2cm}
\subsection{Static state estimation in transmission grids}\label{sec.StaticStateEstimationT}

Consider an electric power network of $n_{bu}$ buses, equipped with $m$ measurements contained in the measurement vector, $\bm{z}$. The measurement model is given by:

\vspace{-.3cm}
\begin{equation}
\bm{z} = \bm{H}\bm{x} + \bm{e}, \label{eq.sysmeasmodel}
\end{equation}

\vspace{-.1cm}
\noindent
where $\bm{H}$ is the measurement matrix, and $\bm{x}$ is the true algebraic state vector. The elements of the measurement error vector, $\bm{e}$, are assumed to be zero-mean, independent, and identically distributed random processes following a Gaussian probability density function, $\bm{e}\sim\mathcal{N}(\bm{0},\bm{R})$---that is, $\mathcal{E}(\bm{e})=\bm{0}$, $\mathcal{E}(\bm{e}\bm{e}\tran )=\bm{R}=\text{diag}(\sigma_{1}^{2},...,\sigma_{m}^{2})$, where $\mathcal{E}$ denotes the expectation operator, $\bm{R}$ is the error covariance matrix, and $\sigma_{k}$ is the standard deviation associated with $k$-th measurement $z_{k}$. This is a convenient assumption because the widely used weighted least squares (WLS) estimator [see the box \emph{Weighted least squares (WLS) estimator}] is the maximum likelihood estimator under these conditions. Note that the measurement model (\ref{eq.sysmeasmodel}) is linear because PMUs measure both the magnitude and phase angle of voltages and currents \cite{Phadke1986}. For example, suppose that buses $\{2,4,6,7\}$ in the network shown in Fig. \ref{fig.7bus} are equipped with PMUs. In this case, the measurement model is given explicitly in (\ref{eq.sysmeasmodel2}), where $v_{k}=|v_{k}|e^{j\theta_{k}}$, $|v_{k}|$ ($\theta_{k}$) is the voltage magnitude (phase angle) at bus $k$; $\tilde{v}_{k}$ ($v_{k}$) denotes the measured (true) algebraic state variable associated with bus $k$; $i_{k\ell}=|i_{k\ell}|e^{j\alpha_{k\ell}}$, $|i_{k\ell}|$ ($\alpha_{k\ell}$) is the current magnitude (phase angle) from bus $k$ to $\ell$; $\tilde{i}_{k\ell}$ denotes the measured current phasor from bus $k$ to $\ell$; $y_{k0}$ is the shunt admittance of bus $k$; and $y_{k\ell}$ is the series admittance between buses $k$ and $\ell$. Hereafter, unless otherwise stated, it is assumed that a PMU installed on bus $k$ measures the voltage phasor at bus $k$ as well as the current phasor in all the branches that are incident to bus $k$. For example, if a PMU is placed on Bus 2 of the network in Fig. \ref{fig.7bus}, then measurements of $\{v_{2},i_{21},i_{23},i_{26},i_{27}\}$ are supposedly available from this device. In other words, the number of measurement channels in each PMU is assumed to be unlimited unless otherwise stated.

\begin{tcolorbox}[colback=black!5!white,
colframe=white!20!black,
title=\textsc{Weighted least squares (WLS) estimator},
center, 
valign = center, 
halign = justify,
before skip = 0.1cm, 
after skip = 0.1cm,
center title, 
width = 1.0\linewidth,
left = 0.1cm,
right = 0.1cm,
before upper = {\parindent1em},
floatplacement = t, float]

A state estimate, $\widehat{\bm{x}}$, is obtained by minimizing an objective function of choice, $J$, as follows:

\vspace{-.3cm}
\begin{equation}
\widehat{\bm{x}} = \argmin_{\bm{x}} \quad J(\bm{x}).
\end{equation}

\vspace{-.1cm}
The WLS estimator minimizes a quadratic criterion:

\vspace{-.3cm}
\begin{equation}
J(\bm{x}) = \frac{1}{2}\left(\bm{z}-\bm{H}\bm{x}\right)\tran \bm{R}^{-1}\left(\bm{z}-\bm{H}\bm{x}\right).
\end{equation}

\vspace{-.1cm}
The solution, $\widehat{\bm{x}}$, is obtained by setting to zero the partial derivative of $J(\bm{x})$ with respect to $\bm{x}$:

\vspace{-.5cm}
\begin{equation}\label{eq.23b}
\widehat{\bm{x}} = \left. \frac{\partial J(\bm{x})}{\partial \bm{x}} \right|_{\bm{x}=\widehat{\bm{x}}} = -\bm{H}\tran \bm{R}^{-1}\left(\bm{z}-\bm{H}\widehat{\bm{x}}\right) = -\bm{H}\tran \bm{R}^{-1}\bm{r} = \bm{0},
\end{equation}

\vspace{-.2cm}
\noindent
where the residual vector $\bm{r} = \bm{z}-\bm{H}\widehat{\bm{x}}$. From (\ref{eq.23b}), we have:

\vspace{-.2cm}
\begin{equation}
-\bm{H}\tran \bm{R}^{-1}\bm{z} + \bm{H}\tran \bm{R}^{-1}\bm{H}\widehat{\bm{x}} = \bm{0},
\end{equation}

\vspace{-.3cm}
\noindent
yielding to:

\vspace{-.4cm}
\begin{equation}\label{eq.24c}
\widehat{\bm{x}} = \left(\bm{H}\tran \bm{R}^{-1}\bm{H}\right)^{-1}\bm{H}\tran \bm{R}^{-1}\bm{z} = \bm{G}^{-1}\bm{H}\tran \bm{R}^{-1}\bm{z}.
\end{equation}

\vspace{-.1cm}
From (\ref{eq.24c}), if the gain matrix, $\bm{G}$, is nonsingular, then (\ref{eq.sysmeasmodel}) is numerically observable. Recall that $\bm{R}$ is a diagonal matrix.
\end{tcolorbox}

\begin{figure}[!t]
\centering
\begin{tikzpicture}[line cap=round,line join=round,>=triangle 45,x=1cm,y=1cm,scale=0.9]
\clip(7.3,3.1) rectangle (13.2,6.1);
\draw [line width=1.0pt] (7.6,5.2)-- (7.6,4);
\draw [line width=1.0pt] (8.8,5.2)-- (8.8,4);
\draw [line width=1.0pt] (12.8,5.2)-- (12.8,4);
\draw [line width=1.0pt] (10.4,5.2)-- (10.4,4);
\draw [line width=1.0pt] (11.6,5.2)-- (11.6,4);
\draw [line width=1.0pt] (9,6)-- (10.2,6);
\draw [line width=1.0pt] (9.8,3.2)-- (11,3.2);
\draw [line width=0.4pt] (7.6,4.6)-- (12.8,4.6);
\draw [line width=0.4pt] (9.8,6)-- (9.8,5);
\draw [line width=0.4pt] (9.8,5)-- (10.4,5);
\draw [line width=0.4pt] (8.8,5)-- (9.4,5);
\draw [line width=0.4pt] (9.4,5)-- (9.4,6);
\draw [line width=0.4pt] (8.8,4.2)-- (10,4.2);
\draw [line width=0.4pt] (10,4.2)-- (10,3.2);
\draw [line width=0.4pt] (10.8,3.2)-- (10.8,4.2);
\draw [line width=0.4pt] (10.8,4.2)-- (11.6,4.2);
\draw ( 7.20,4.25) node[anchor=north west] {1};
\draw ( 8.40,4.25) node[anchor=north west] {2};
\draw (10.20,6.15) node[anchor=north west] {6};
\draw (11.65,6.15) node[anchor=north west] {PMU};
\draw (10.40,5.25) node[anchor=north west] {3};
\draw (11.60,4.25) node[anchor=north west] {4};
\draw (12.80,4.25) node[anchor=north west] {5};
\draw ( 9.40,3.45) node[anchor=north west] {7};
\draw ( 8.00,5.00) node[anchor=north west] {i};
\draw ( 9.30,5.00) node[anchor=north west] {ii};
\draw ( 8.95,5.70) node[anchor=north west] {iii};
\draw ( 9.55,4.00) node[anchor=north west] {iv};
\draw (10.85,4.95) node[anchor=north west] {v};
\draw ( 9.75,5.70) node[anchor=north west] {vi};
\draw (11.95,5.00) node[anchor=north west] {vii};
\draw (10.75,4.00) node[anchor=north west] {viii};
\begin{scriptsize}
\draw [fill=red] (9.2,6) circle (2.0pt);
\draw [fill=red] (8.8,4.8) circle (2.0pt);
\draw [fill=red] (10.4,3.2) circle (2.0pt);
\draw [fill=red] (11.6,4.8) circle (2.0pt);
\draw [fill=red] (11.6,5.9) circle (2.0pt);
\end{scriptsize}
\end{tikzpicture}
\caption{One-line diagram of the 7-bus system \cite{Xu2004}. The bus (branch) indices are indicated by Arabic (Roman) numerals.}
\label{fig.7bus}
\end{figure}
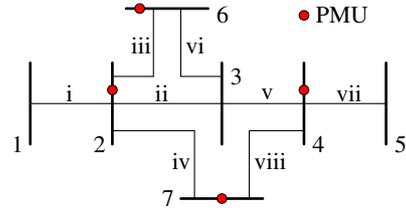

\begin{figure*}[!ht]
\begin{equation}\label{eq.sysmeasmodel2}
\left[{\def\arraystretch{0.8}
\begin{array}{c}
\tilde{v}_{2} \\
\tilde{v}_{4} \\
\tilde{v}_{6} \\
\tilde{v}_{7} \\
\tilde{i}_{21} \\
\tilde{i}_{23} \\
\tilde{i}_{26} \\
\tilde{i}_{27} \\
\tilde{i}_{43} \\
\tilde{i}_{45} \\
\tilde{i}_{47} \\
\tilde{i}_{62} \\
\tilde{i}_{63} \\
\tilde{i}_{72} \\
\tilde{i}_{74}
\end{array}}
\right]=
\left[{\def\arraystretch{1}
\begin{array}{r r r r r r r}
0 & 1 & 0 & 0 & 0 & 0 & 0 \\
0 & 0 & 0 & 1 & 0 & 0 & 0 \\
0 & 0 & 0 & 0 & 0 & 1 & 0 \\
0 & 0 & 0 & 0 & 0 & 0 & 1 \\
-y_{12} & y_{20}+y_{12} & 0 & 0 & 0 & 0 & 0 \\
0 & y_{20}+y_{23} & -y_{23} & 0 & 0 & 0 & 0 \\
0 & y_{20}+y_{26} & 0 & 0 & 0 & -y_{26} & 0 \\
0 & y_{20}+y_{27} & 0 & 0 & 0 & 0 & -y_{27} \\
0 & 0 & -y_{34} & y_{40}+y_{34} & 0 & 0 & 0 \\
0 & 0 & 0 & y_{40}+y_{45} & -y_{45} & 0 & 0 \\
0 & 0 & 0 & y_{40}+y_{47} & 0 & 0 & -y_{47} \\
0 & -y_{26} & 0 & 0 & 0 & y_{60}+y_{26} & 0 \\
0 & 0 & -y_{36} & 0 & 0 & y_{60}+y_{36} & 0 \\
0 & -y_{27} & 0 & 0 & 0 & 0 & y_{70}+y_{27} \\
0 & 0 & 0 & -y_{47} & 0 & 0 & y_{70}+y_{47}
\end{array}}
\right]
\left[{\def\arraystretch{0.8}
\begin{array}{c}
v_{1} \\
v_{2} \\
v_{3} \\
v_{4} \\
v_{5} \\
v_{6} \\
v_{7}
\end{array}}
\right]+
\left[{\def\arraystretch{1}
\begin{array}{c}
e_{1} \\
e_{2} \\
e_{3} \\
e_{4} \\
e_{5} \\
e_{6} \\
e_{7} \\
e_{8} \\
e_{9} \\
e_{10} \\
e_{11} \\
e_{12} \\
e_{13} \\
e_{14} \\
e_{15}
\end{array}}
\right].
\end{equation}
\noindent\makebox[\linewidth]{\rule{\textwidth}{0.4pt}}
\end{figure*}

\begin{tcolorbox}[colback=black!5!white,
colframe=white!20!black,
title=\textsc{Observability for static state estimation in transmission grids}, 
center, 
valign = center, 
halign = justify,
before skip = 0.1cm, 
after skip = 0.1cm,
center title, 
width = 1.0\linewidth,
left = 0.1cm,
right = 0.1cm,
before upper = {\parindent1em},
floatplacement = t, float]
The key contributions to power system observability analysis are attributed to Clements, Krumpholz, and Davis \cite{Krumpholz1980, Clements1982}. Two definitions of observability are well accepted:

\vspace{.1cm}
\textbf{Numerical observability} is defined as the ability of the measurement model (\ref{eq.sysmeasmodel}) to be solved for a state estimate, $\widehat{\bm{x}}$. If $\bm{H}$ in (\ref{eq.sysmeasmodel}) is of full rank and well conditioned, or, equivalently, if $\bm{G}$ in (\ref{eq.24c}) is nonsingular, then the system is said to be numerically observable.

\vspace{.1cm}
\textbf{Topological observability} is defined as the existence of at least one spanning measurement tree of full rank in the network. 

\vspace{.1cm}
Numerical observability implies topological observability, but the converse is not true. In practice, however, cases where a power grid is topologically but not numerically observable are rare. See, e.g., \cite{Clements1990} for more details.
\end{tcolorbox}

\subsubsection{Observability for static state estimation in transmission grids}\label{subsec.ObservabilityT}$\,$

\vspace{.1cm}
\noindent
\emph{Minimum PMU placement for observability in ideal conditions:}

\vspace{.1cm}
The most basic requirement for state estimation is the observability of the measurement model (\ref{eq.sysmeasmodel}) [see the box \emph{Observability for static state estimation in transmission grids}]. This is because $\bm{H}$ is assumed to be perfectly known; thus, if (\ref{eq.sysmeasmodel}) is observable, one can rely on the measurements in $\bm{z}$ to obtain an estimate, $\widehat{\bm{x}}$, of the algebraic state vector. On the other hand, installing a PMU on each bus of an electric transmission grid is cost-prohibitive; hence, a relevant question to ask is, what is the minimum set of PMUs that makes the measurement model (\ref{eq.sysmeasmodel}) observable? This question is investigated in \cite{Mili1990, Baldwin1993} by extending the notion of a spanning measurement tree to the specific case of PMUs. Following \cite{Baldwin1993}, a spanning measurement tree is any network subgraph that \emph{``contains all the nodes of the network and has an actual measurement or a calculated pseudo-measurement assigned to each of its branches. A pseudo-measurement is assigned to either a non-metered branch where the voltage phasor at both ends are known [using Ohm's law], or to a non-metered branch which is incident to a bus where all but the current of that branch are known [using Kirchhoff's current law].''} In particular, it is shown in \cite{Baldwin1993} through numerical simulations performed on various test systems that approximately:

\vspace{-0.3cm}
\begin{itemize}
\item One-fourth to one-third of the network buses \emph{in general} and
\item One-half of the network buses \emph{in the worst-case scenario}
\end{itemize}

\vspace{-0.3cm}
\noindent
need to be instrumented with PMUs to achieve observability. The pioneering work in \cite{Mili1990, Baldwin1993} sets the ground for the minimum PMU placement problem in transmission grids. It relies on a dual search algorithm, comprising a modified bisecting search and a simulated annealing method, to build the spanning measurement tree of the network. When computing time is of concern, several modifications can be applied to the heuristics in \cite{Baldwin1993} to accelerate the solution process \cite{Cho2001}. See also \cite{Denegri2002} for some variations of the heuristics in \cite{Baldwin1993} and \cite{Peng2006} for a method based on Tabu search. 

The reader interested in a formal treatment of the minimum PMU placement problem is referred to \cite{Brueni1993, Brueni2005}, where a graph-theoretic approach is taken to prove that, under an ideal scenario, no more than one-third of the network buses need to be provided with PMUs for observability---we stress that this holds only under ideal scenarios. In fact, later in this section, we discuss a counterexample where more than one-half of the network buses need to be instrumented with PMUs to achieve observability. It is also shown in \cite{Brueni1993, Brueni2005} that the minimum PMU placement problem is NP-complete. The metaheuristics used to find a solution to the problem under discussion are summarized in Table \ref{tab.2}.

\begin{table}[htb!]
\centering \scriptsize
\setlength{\tabcolsep}{0.2em}
\caption{Metaheuristics used to find the minimum set of PMUs for observability}
\begin{tabular}{l l l l}
\cmidrule[1pt]{1-4}
\rowcolor{black!10}\textbf{Metaheuristic} & \textbf{Ref.} & \textbf{Metaheuristic} & \textbf{Ref.} \\ \hline
Bisecting search & \cite{Mili1990, Baldwin1993} & Fuzzy logic & \cite{Sodhi2015} \\
Cellular learning automata & \cite{Mazhari2013} & Genetic algorithm & \cite{Marin2003, Milosevic2003, deSouza2005, Aminifar2009, Shahraeini2012, Muller2016} \\
Chemical reaction optimization & \cite{Wen2013} & Iterated local search & \cite{Hurtgen2010} \\
Cuckoo search algorithm & \cite{Dalali2016} & Particle swarm optimization & \cite{Hajian2011, Ahmadi2011, Rahman2017, Rather2015} \\
Evolutionary algorithm & \cite{Peng2010, Jamuna2012, Mohammadi2016} & Simulated annealing & \cite{Mili1990, Baldwin1993, Cho2001, Nuqui2002, Nuqui2005} \\
Exhaustive binary search & \cite{Chakrabarti2008} & Tabu search & \cite{Cho2001, Peng2006, Koutsoukis2013} \\ \hline
\cmidrule[1pt]{1-4}
\end{tabular}
\label{tab.2}
\end{table}

Note, reference \cite{Chakrabarti2008} in Table \ref{tab.2} uses an exhaustive binary search to find the minimum number of PMUs that makes (\ref{eq.sysmeasmodel}) topologically observable. Despite being computationally expensive, the exhaustive search provides a globally optimal solution; thus, the results obtained for the systems in \cite{Chakrabarti2008}---IEEE 14-bus, IEEE 24-bus, IEEE 30-bus, and New England 39-bus---serve as a benchmark solution. Other than an exhaustive search, the metaheuristics in Table \ref{tab.2} do not guarantee a globally optimal solution. Unfortunately, an exhaustive search becomes quickly impractical as the number of buses (and thereby the dimension of the problem) increases because the problem is NP-complete.

Instead of a heuristic, the formulation in \cite{Xu2004} relies on an optimization model, as follows:

\vspace{-.4cm}
\begin{equation}\label{eq.2}
\begin{aligned}
\min_{} \quad & \sum_{k=1}^{n_{bu}}c_{k}\cdot u_{k}, \\
\textrm{s.t.} \quad & \bm{g}(\bm{u}) \ge \bm{1}_{n_{bu}},
\end{aligned}
\end{equation}

\noindent
where $c_{k}$ denotes the cost of installing a PMU on bus $k$; $g_{k}(\bm{u})\in\mathbb{Z}$, $\bm{1}_{n_{bu}}$ denotes the vector of all ones of dimension $n_{bu}$; and:

\vspace{-.2cm}
\begin{equation}\label{eq.4b}
u_{k}=
\begin{cases}
1 \quad \text{if a PMU is installed on bus } k,\\
0 \quad \text{otherwise}.
\end{cases}
\end{equation}

\vspace{-.1cm}
The solution to the optimization problem in (\ref{eq.2}) is obtained by integer programming (IP). For this reason, (\ref{eq.2}) is hereafter referred to as the \emph{IP method}. During the solution process, the inequality constraints associated with $\bm{g}(\bm{u})$ enforce topological observability. The rationale of the mechanism by which topological observability is enforced is explained next. Consider:

\vspace{-.4cm}
\begin{equation}\label{eq.4a}
g_{k}(\bm{u})=\sum_{\ell=1}^{n_{bu}}A_{k\ell} \cdot u_{\ell}\ge 1,
\end{equation}

\vspace{-.2cm}
\noindent
where:

\vspace{-.6cm}
\begin{equation}\label{eq.5a}
A_{k\ell}=
\begin{cases}
1 \quad \text{if } k=\ell \text{, or if buses } k \text{ and } \ell \text{ are connected},\\
0 \quad \text{otherwise},
\end{cases}
\end{equation}

\vspace{-.3cm}
\noindent
denotes the $k\ell$-th element of the bus-to-bus connectivity matrix [see the box \emph{Admittance matrix and connectivity matrices}]. The constraints defined in (\ref{eq.4a}) are better explained through an example. To this end, consider again the network in Fig. \ref{fig.7bus}. From (\ref{eq.4a})--(\ref{eq.5a}), we have:

\vspace{-.5cm}
\begin{equation}\label{eq.7a}
\bm{A}=\left[{\arraycolsep=1.8pt\def\arraystretch{1.1}
\begin{array}{c c c c c c c}
1 & 1 & 0 & 0 & 0 & 0 & 0 \\
1 & 1 & 1 & 0 & 0 & 1 & 1 \\
0 & 1 & 1 & 1 & 0 & 1 & 0 \\
0 & 0 & 1 & 1 & 1 & 0 & 1 \\
0 & 0 & 0 & 1 & 1 & 0 & 0 \\
0 & 1 & 1 & 0 & 0 & 1 & 0 \\
0 & 1 & 0 & 1 & 0 & 0 & 1 
\end{array}}
\right], \;
\text{and }
\begin{cases}
\begin{array}{l}
g_{1} = u_{1} + u_{2} \ge 1, \\
g_{2} = u_{1} + u_{2} + u_{3} + u_{6} + u_{7} \ge 1, \\
g_{3} = u_{2} + u_{3} + u_{4} + u_{6} \ge 1, \\
g_{4} = u_{3} + u_{4} + u_{5} + u_{7} \ge 1, \\
g_{5} = u_{4} + u_{5} \ge 1, \\
g_{6} = u_{2} + u_{3} + u_{6} \ge 1, \\
g_{7} = u_{2} + u_{4} + u_{7} \ge 1. 
\end{array}
\end{cases}
\end{equation}

\begin{tcolorbox}[colback=black!5!white,
colframe=white!20!black,
title=\textsc{Admittance matrix and connectivity matrices},
center, 
valign = center, 
halign = justify,
before skip = 0.1cm, 
after skip = 0.1cm,
center title, 
width = 1.0\linewidth,
left = 0.1cm,
right = 0.1cm,
before upper = {\parindent1em},
floatplacement = t, float]

The \textbf{admittance matrix}---also called $\bm{Y}_{bus}$ or $\bm{Y}\,\text{\emph{matrix}}$ or \emph{bus admittance matrix} or \emph{nodal admittance matrix}---is a matrix of dimension $n_{bu} \times n_{bu}$ that represents the nodal admittance of electric power networks. The $\bm{Y}_{bus}$ is likely the most used matrix in power system studies, and it has several interesting properties. For example, the $\bm{Y}_{bus}$ is:

\vspace{-0.3cm}
\begin{itemize}
\item Very sparse for real electric power transmission grids;
\item Symmetric if no phase-shifting transformers are included.
\end{itemize}

\vspace{-0.3cm}
For the network shown in Fig. \ref{fig.7bus}:

\vspace{-.4cm}
\begin{equation*}
\bm{Y}=\left[{\arraycolsep=3.0pt\def\arraystretch{1.0}
\begin{array}{r r r r r r r}
y_{11} & -y_{12} & 0 & 0 & 0 & 0 & 0 \\
-y_{12} & y_{22} & -y_{23} & 0 & 0 & -y_{26} & -y_{27} \\
0 & -y_{23} & y_{33} & -y_{34} & 0 & -y_{36} & 0 \\
0 & 0 & -y_{34} & y_{44} & -y_{45} & 0 & -y_{47} \\
0 & 0 & 0 & -y_{45} & y_{55} & 0 & 0 \\
0 & -y_{26} & -y_{36} & 0 & 0 & y_{66} & 0 \\
0 & -y_{27} & 0 & -y_{47} & 0 & 0 & y_{77} 
\end{array}}
\right].
\end{equation*}
 
\vspace{-0.2cm}
One can use $\bm{Y}$, which is readily available, to build the \textbf{bus-to-bus connectivity matrix}, $\bm{A}$. To do so, simply set all nonzero elements of $\bm{Y}$ to $1$. Also, $\bm{A}$ can be built by using (\ref{eq.5a}).

\vspace{.1cm}
The \textbf{bus-to-branch connectivity matrix}, denoted by $\bm{B}$ in (\ref{eq.19a}), can be built by using (\ref{eq.18a}). See, e.g., \cite{Zimmerman2011} for more details.
\end{tcolorbox}

The constraints in (\ref{eq.7a}) guarantee topological observability by enforcing that at least one $u_{\ell}=1$ for each $g_{k}$, $k,\ell=(1,...,n_{bu})$. Or, in plain words, they guarantee that the voltage phasor on each bus in the network is either directly measured or indirectly calculated. The IP method is well accepted because of its simplicity and scalability. The reader seeking an in-depth understanding should start with \cite{Chen1990} and the references therein. The IP method can be modified to consider measurements acquired from RTUs and/or pseudo-measurements inferred from zero-injection buses. This is accomplished by modifying the constraints associated with $\bm{g}(\bm{u})$. Note that a zero-injection bus is a bus without generators, loads, or any other shunt device connected to it. For example, let Bus 3 in the 7-bus system shown in Fig. \ref{fig.7bus} be a zero-injection bus. If the voltage phasor in any three buses in the set $\{2,3,4,6\}$ is measured, then the voltage phasor at the fourth bus can be calculated by applying Kirchhoff's current law at Bus 3, where the net injected current is known. Following \cite{Xu2004}, the constraints in (\ref{eq.7a}) are modified as follows:

\vspace{-.2cm}
\begin{equation}\label{eq.13z}
\begin{array}{l}
g_{2} = u_{1} + u_{2} + u_{3} + u_{6} + u_{7} + g_{3}\cdot g_{4}\cdot g_{6} \ge 1, \\
g_{4} = u_{3} + u_{4} + u_{5} + u_{7} + g_{2}\cdot g_{3}\cdot g_{6} \ge 1, \\
g_{6} = u_{2} + u_{3} + u_{6} + g_{2}\cdot g_{3}\cdot g_{4} \ge 1.
\end{array}
\end{equation}

By replacing $g_{2}$, $g_{3}$, $g_{4}$, and $g_{6}$ in (\ref{eq.13z}) and simplifying via Boolean logic, one obtains:

\vspace{-.3cm}
\begin{equation}\label{eq.14z}
\begin{array}{l}
g_{2} = u_{1} + u_{2} + u_{3} + u_{6} + u_{7} \ge 1, \\
g_{4} = u_{2} + u_{3} + u_{4} + u_{5} + u_{6} + u_{7} \ge 1, \\
g_{6} = u_{2} + u_{3} + u_{6} + u_{1} \cdot u_{4} + u_{4} \cdot u_{7} \ge 1.
\end{array}
\end{equation}

Note that the modification suggested in \cite{Xu2004} makes the constraints in (\ref{eq.14z}) nonlinear, which is not desirable. A better approach to modify the constraints in (\ref{eq.2}) is developed in \cite{Gou2008a, Gou2008b} and extended in \cite{Gou2014}. The interested reader is referred to those papers for details. See also \cite{Abbasy2009, Aminifar2010, Kavasseri2011, Azizi2012, Esmaili2013}. Note that considering zero-injection buses is desirable because it reduces the number of PMUs required for system observability. The IP method can also be modified to account for the single loss of any PMU \cite{Xu2005} and other relevant factors. Extensions of the IP method will be discussed in several opportunities throughout the paper. The trade-offs between some of these extensions and the obtained solutions are well presented in \cite{Aminifar2010, Khajeh2017}. Specifically, in \cite{Aminifar2010}, the set of constraints in the IP method is extended such that:

\vspace{-.3cm}
\begin{itemize}
\item Zero-injection buses are considered for PMU installation.
\item The system remains observable in case of the loss of any single PMU.
\item The system remains observable in case of single branch outages.
\item The number of measurement channels in some or all PMUs is limited.
\end{itemize}

\vspace{-.3cm}
Moreover, \cite{Aminifar2010} suggests a specific optimization solver in which an optimality gap can be specified as a trade-off between a slower/optimal and a faster/suboptimal solution.

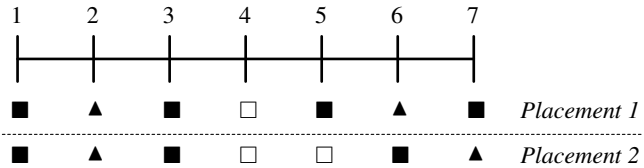
\begin{figure}[H]
\centering
\definecolor{red}{rgb}{1,0,0}
\begin{tikzpicture}[line cap=round,line join=round,>=triangle 45,x=1cm,y=1cm]
\clip(7.2,3.55) rectangle (15.7,5.7);
\draw [line width=1.0pt] ( 7.5,5.3)-- ( 7.5,4.7);
\draw [line width=1.0pt] ( 8.5,5.3)-- ( 8.5,4.7);
\draw [line width=1.0pt] ( 9.5,5.3)-- ( 9.5,4.7);
\draw [line width=1.0pt] (10.5,5.3)-- (10.5,4.7);
\draw [line width=1.0pt] (11.5,5.3)-- (11.5,4.7);
\draw [line width=1.0pt] (12.5,5.3)-- (12.5,4.7);
\draw [line width=1.0pt] (13.5,5.3)-- (13.5,4.7);
\draw [line width=1.0pt] ( 7.5,5)-- (13.5,5);
\draw ( 7.30,5.8) node[anchor=north west] {1};
\draw ( 8.30,5.8) node[anchor=north west] {2};
\draw ( 9.30,5.8) node[anchor=north west] {3};
\draw (10.30,5.8) node[anchor=north west] {4};
\draw (11.30,5.8) node[anchor=north west] {5};
\draw (12.30,5.8) node[anchor=north west] {6};
\draw (13.30,5.8) node[anchor=north west] {7};
\draw [line width=0.4pt,dash pattern=on 1pt off 1pt] (7.3,4.0)-- (15.8,4.0);
\draw ( 7.3,4.54) node[anchor=north west] {$\blacksquare$};
\draw ( 8.3,4.54) node[anchor=north west] {$\blacktriangle$};
\draw ( 9.3,4.54) node[anchor=north west] {$\blacksquare$};
\draw (10.3,4.54) node[anchor=north west] {$\square$};
\draw (11.3,4.54) node[anchor=north west] {$\blacksquare$};
\draw (12.3,4.54) node[anchor=north west] {$\blacktriangle$};
\draw (13.3,4.54) node[anchor=north west] {$\blacksquare$};
\draw (14.0,4.55) node[anchor=north west] {\emph{Placement 1}};
\draw ( 7.3,3.94) node[anchor=north west] {$\blacksquare$};
\draw ( 8.3,3.94) node[anchor=north west] {$\blacktriangle$};
\draw ( 9.3,3.94) node[anchor=north west] {$\blacksquare$};
\draw (10.3,3.94) node[anchor=north west] {$\square$};
\draw (11.3,3.94) node[anchor=north west] {$\square$};
\draw (12.3,3.94) node[anchor=north west] {$\blacksquare$};
\draw (13.3,3.94) node[anchor=north west] {$\blacktriangle$};
\draw (14.0,3.95) node[anchor=north west] {\emph{Placement 2}};
\end{tikzpicture}
\caption{The symbol $\blacktriangle$ indicates the bus where the PMU is placed and thus its voltage is directly measured. $\blacksquare$ indicates that the voltage at that bus can be calculated. $\square$ indicates that the voltage at that bus is not accessible.}
\label{fig.degree_of_unobservability}
\end{figure}

In practice, the deployment of, e.g., $n_{bu}/3$ PMUs, requires an enormous effort, even for limited-size networks. Considering this practical aspect, the work in \cite{Nuqui2002, Nuqui2005} develops the concept of \emph{degree of unobservability}, which is better explained through an example. Consider the network in Fig. \ref{fig.degree_of_unobservability}. In Placement 1, the following quantities are measured: $\{v_{2},i_{21},i_{23},v_{6},i_{65},i_{67}\}$; thus, the voltage phasor at buses $\{2,6\}$ is directly measured by PMUs, and the voltage phasor at buses $\{1,3,5,7\}$ can be calculated by using the following relation:

\vspace{-.4cm}
\begin{equation}\label{eq.8b}
\left[{\arraycolsep=0.6pt\def\arraystretch{0.9}
\begin{array}{c}
i_{21} \\
i_{23} \\
i_{65} \\
i_{67}
\end{array}}
\right] =
\left[{\arraycolsep=0.8pt\def\arraystretch{0.8}
\begin{array}{rrrrrr}
y_{20}+y_{12} & 0 & -y_{12} & 0 & 0 & 0 \\
y_{20}+y_{23} & 0 & 0 & -y_{23} & 0 & 0 \\
0 & y_{60}+y_{56} & 0 & 0 & -y_{56} & 0 \\
0 & y_{60}+y_{67} & 0 & 0 & 0 & -y_{67}
\end{array}}
\right]
\left[{\arraycolsep=0.6pt\def\arraystretch{0.9}
\begin{array}{c}
v_{1} \\
v_{3} \\
v_{5} \\
v_{7} \\
\end{array}}
\right].
\end{equation}

\vspace{-.6cm}

\vspace{-.1cm}
Conversely, $v_{4}$ cannot be calculated unless Bus 4 is a zero-injection bus \cite{Cho2001}. The system with the measurement Placement 1 is said to be a system with a \emph{depth-of-one unobservability} because there is at least one bus in which the voltage is not accessible, linked to two or more buses in which the voltage is calculated. Similarly, Placement 2 defines a system with a \emph{depth-of-two unobservability}. It is stated in \cite{Nuqui2005} that \emph{``the crux of this approach is that the state [voltage phasor] of buses that are not observable can be interpolated from the state of their neighbors rather accurately,''} although the level of accuracy is not discussed. The idea is to use the degree of unobservability to strategically plan for a phased, or multistage, installation of PMUs. The work in \cite{Pal2014} fuses the concept of the degree of unobservability with the IP method and develops a placement scheme that prioritizes the real-time monitoring of \emph{critical} buses of the network. Buses are defined to be critical if they satisfy one or more of the following conditions:

\vspace{-.3cm}
\begin{itemize}
\item Are high voltage, i.e., 200 kV and more
\item Have many incident branches
\item Are relevant to rotor angle stability
\item Are relevant to small-signal stability,
\end{itemize}

\vspace{-.3cm}
\noindent
and they are selected based on stability assessments performed offline. The PMU placement problem is formulated to ensure that the voltage phasor of critical buses is directly or indirectly accessible by at least one PMU, with weights assigned to prioritize the critical buses. 

The weakness of relying solely on the concept of degree of unobservability to allocate PMUs in a multistage fashion is that upon completion of the last stage, the number of installed PMUs will be larger than the minimum necessary for observability. In other words, this approach leads to a larger number of placed PMUs than \cite{Baldwin1993}, \cite{Xu2004}, and others. Nonetheless, the idea of installing PMUs in a phased manner is of practical interest given the amount of investment involved. Along these lines, let $\mathbb{S}$, $\text{card}(\mathbb{S})=n_{p}$, be the minimum set of PMUs required to attain observability. Ideally, $\text{card}(\mathbb{S})=\text{card}(\mathbb{S}_{1}\cup\mathbb{S}_{2}\cup\cdots\cup\mathbb{S}_{t})$, where $\mathbb{S}_{t}$ denotes the set of PMUs installed in the last stage, $t$. This condition would ensure that upon completion of the last stage, the number of PMUs installed is not larger than the minimum set of PMUs required to attain observability. On the other hand, the measurement model will not be observable until the last stage is completed. This idea is pursued in \cite{Dua2008}, which accounts for an important extension of the IP method. The minimum set of PMUs required to attain observability is determined beforehand. Also, the number of PMUs to be installed at each stage is defined a priori based on, e.g., the available budget for each stage. Then, for each stage $\ell$:

\vspace{-.3cm}
\begin{equation}\label{eq.9c}
\addtolength{\jot}{-.4em}
\begin{aligned}
\max_{} \quad & \sum_{k=1}^{n_{bu}}s_{k}, \\
\textrm{s.t.} \quad & \bm{A}\bm{u} \ge \bm{s}, \\
& u_{k}=
\begin{cases}
0 \quad\forall\, k \not\in\mathbb{S}, \\
1 \quad\forall\, k \in\mathbb{S}_{0}\cup\mathbb{S}_{1}\cup\cdots\cup\mathbb{S}_{t-1},
\end{cases} \\
& \sum_{k=1}^{n_{bu}}u_{k} = \sum_{\ell=1}^{t}n_{\ell}. \\
\end{aligned}
\end{equation}

\vspace{-.1cm}
The interpretation of (\ref{eq.9c}) is simple. The set $\mathbb{S}$ is known, but not all PMUs can be installed at once. The question is what is the best choice of $\mathbb{S}_{\ell}\subset\mathbb{S}$, $\ell=1,...,t$. The strategy in (\ref{eq.9c}) is, for each stage $\ell$, to maximize the number of buses in which the voltage phasor is measured or can be calculated, given the number of PMUs to be installed at stage $\ell$, $n_{\ell}<n_{p}$. In (\ref{eq.9c}), $s_{k}$ is equal to 1 if the voltage at bus $k$ is directly measured or can be calculated, and 0 otherwise. The inequality $\bm{A}\bm{u} \ge \bm{s}$ is a relaxation of (\ref{eq.4a}). The equality $u_{k}=0 \;\forall\, k \not\in\mathbb{S}$ ensures that only elements of $\mathbb{S}$ are candidates for placement, whereas $u_{k}=1 \;\forall\, k \in\mathbb{S}_{0}\cup\mathbb{S}_{1}\cup\cdots\cup\mathbb{S}_{t-1}$ fixes the candidates chosen in earlier stages; $\mathbb{S}_{0}=\varnothing$. The last equality constraint restricts the number of PMUs allowed to be placed at stage $\ell$. The formulation in (\ref{eq.9c}) is amenable to the consideration of zero-injection buses, and the constraints remain linear, as is the case in \cite{Gou2008a, Gou2008b}; therefore, this formulation is henceforth referred to as the \emph{multistage ILP method}. In addition to (\ref{eq.9c}), the work in \cite{Dua2008} develops a simple yet effective way of including a measure of the degree of redundancy in the optimization model. The number of PMUs that provides access to the voltage phasor at bus $k$ is:

\vspace{-.3cm}
\begin{equation}
n_{pmu}^{(k)} = n_{br}^{(k)} + 1,
\end{equation}

\vspace{-.2cm}
\noindent
where $n_{br}^{(k)}$ denotes the number of branches connected to bus $k$. A measure of the degree of redundancy is given by:

\vspace{-.3cm}
\begin{equation}\label{eq.11c}
\gamma = \sum_{k=1}^{n_{bu}} n_{pmu}^{(k)},
\end{equation}

\vspace{-.1cm}
\noindent
and it can be easily incorporated into the IP method and its extensions. Similar ideas are used in, e.g., \cite{London2007, Chakrabarti2009}. Note that the degree of redundancy has an important effect on the performance of the well-known robust state estimators, e.g., the least absolute value (LAV) estimator \cite{Abur2004} and the Schweppe-type generalized maximum likelihood estimator \cite{Mili1996}. The work in \cite{Gol2013} modifies the IP method and develops a systematic approach to place PMU measurements such that the minimum degree of redundancy required by the LAV estimator is achieved. Measurement redundancy is mission-critical for state estimation and will be discussed in more detail later in this section. In addition to (\ref{eq.11c}), other criteria can be used to prioritize certain locations in earlier stages of a multistage plan or to rank multiple solutions of single-stage placements. For example, one can prioritize the observability of voltage control areas and/or important tie-lines; see \cite{Sodhi2011}. The multistage minimum PMU placement problem is approached from a probabilistic viewpoint in \cite{Aminifar2011}, where the observability of each bus in a network is modeled by a probability density function. In other words, as opposed to a deterministic view, where each bus in a network is either observable or not, the view in \cite{Aminifar2011} is that each bus in a network is observable with a given probability. Accordingly, the problem constraints are modified, and the IP method is reformulated as a mixed-integer linear programming problem. Unfortunately, the additional complexity added by the probabilistic constraints is exchanged by neglecting the zero-injection buses, which is an important deficiency. This deficiency is addressed in \cite{Wang2014, Aghaei2015}. See also \cite{Wang2020}.

The placement strategies discussed to this point are based on the notion of topological observability, which is not strictly a sufficient condition to solve the state estimation problem (\ref{eq.24c}) \cite{Monticelli1999}. In practice, it is possible to encounter cases in which $\bm{H}$ is numerically ill conditioned. The work in \cite{Madtharad2003} addresses this issue by developing a simple procedure \cite{Slutsker1987} to attain numerical observability [see the box \emph{Observability for static state estimation in transmission grids}]. Note that the dimension of $\bm{H}$ in (\ref{eq.sysmeasmodel}) is $m \times n_{bu}$, and, in general, $m \ge n_{bu}$. In particular, $m=n_{bu}$ is the size of the minimum set of linearly independent measured variables for which (\ref{eq.sysmeasmodel}) can be solved. The idea in \cite{Madtharad2003} is to start by building the matrix $\bm{H}$ considering all possible $n_{bu}$ voltage measurements, plus all possible $n_{br}$ current measurements, such that, initially, $m=n_{bu}+2n_{br}>n_{bu}$. Then, \emph{Algorithm 1} is executed.

\begin{algorithm}[!ht]
\SetAlgoLined
$m \leftarrow$ number of rows of $\bm{H}$\;
$n \leftarrow$ number of columns of $\bm{H}$\;
\While{$m>n$}{
\emph{condNum0} $\leftarrow$ $10^{9}$\;
\For{$k \leftarrow 1$ to $m$}{
$\bm{H}_{aux}$ $\leftarrow$ $\bm{H}$\;
Remove the $k$-th row of $\bm{H}_{aux}$\;
\emph{condNum} $\leftarrow$ condition number of $\bm{H}_{aux}$\;
\If{condNum < condNum0}{
\emph{condNum0} $\leftarrow$ condNum\;
$\ell$ $\leftarrow$ $k$\;
}
}
Remove the $\ell$-th row of $\bm{H}$\;
$m \leftarrow$ $m-1$\;
}
\caption{Incremental reduction of $\bm{H}$}
\end{algorithm}

The rows of $\bm{H}$ that last upon the completion of Algorithm 1 indicate the minimum set of independent variables that guarantees numerical observability. Note that this method leads to the minimum set of independent variables, which does not necessarily correspond to the minimum number of PMUs. This important drawback is alleviated by a heuristic given in \cite{Rakpenthai2005}, but there is no guarantee that the obtained set of PMUs is minimum. This drawback is effectively overcome in \cite{Sodhi2010} by combining the IP method with the original ideas in \cite{Madtharad2003}. The work in \cite{Koutsoukis2013} relies on similar ideas, i.e.: i) build $\bm{H}$ for an initial PMU placement; ii) remove a PMU and update $\bm{H}$; iii) check if the updated $\bm{H}$ is of full rank. The solution is obtained iteratively.

\vspace{.2cm}
\noindent
\emph{Consideration of PMUs with limited measurement channels:}

\vspace{.1cm}
To this point, it is assumed that a PMU installed on bus $k$ measures the voltage phasor at bus $k$ as well as the current phasor in all the branches that are incident to bus $k$. This is possible because of the capability of the commercially available PMUs that typically have more than 20 measurement channels; however, this capability could be limited, thereby influencing the problem solution. To this end, \cite{Emami2008, Emami2010} use a strict assumption that PMUs are supposedly provided with two measurement channels---one for the voltage signal and the other for the current signal. Moreover, though previous work supposes that PMUs are installed on the network \emph{buses}, here, PMUs are supposedly installed on the network \emph{branches}. Accordingly, the optimization problem in (\ref{eq.2})--(\ref{eq.5a}) is reformulated as follows:

\vspace{-.3cm}
\begin{equation}\label{eq.13a}
\begin{aligned}
\min_{} \quad & \sum_{k=1}^{n_{br}}c_{k}^{\prime} \cdot u_{k}^{\prime}, \\
\textrm{s.t.} \quad & \bm{g}(\bm{u}^{\prime}) \ge \bm{1},
\end{aligned}
\end{equation}

\vspace{-.2cm}
\noindent
where $n_{br}$ is the number of branches in the network; $c_{k}^{\prime}$ is the cost of installing a PMU on branch $k$:

\vspace{-.6cm}
\begin{align}
u_{k}^{\prime} &=
\begin{cases}
1 \quad \text{if a PMU is installed on branch } k,\\
0 \quad \text{otherwise},
\end{cases} \\
g_{k}(\bm{u}^{\prime}) &= \sum_{\ell=1}^{n_{br}}B_{k\ell} \cdot u_{\ell}^{\prime}\ge 1, \\
B_{k\ell} &=
\begin{cases}
1 \quad \text{if branch } \ell \text{ is incident to bus } k,\\
0 \quad \text{otherwise},
\end{cases}\label{eq.18a}
\end{align}

\vspace{-.2cm}
\noindent
and $B_{k\ell}$ denotes the $k\ell$-th element of the \emph{bus-to-branch} connectivity matrix $\bm{B}$ [see the box \emph{Admittance matrix and connectivity matrices}]. For example, for the network in Fig. \ref{fig.7bus}:

\vspace{-.2cm}
\begin{equation}\label{eq.19a}
\bm{B} = \left[{\arraycolsep=3.0pt\def\arraystretch{0.8}
\begin{array}{c c c c c c c c}
1 & 0 & 0 & 0 & 0 & 0 & 0 & 0 \\
1 & 1 & 1 & 1 & 0 & 0 & 0 & 0 \\
0 & 1 & 0 & 0 & 1 & 1 & 0 & 0 \\
0 & 0 & 0 & 0 & 1 & 0 & 1 & 1 \\
0 & 0 & 0 & 0 & 0 & 0 & 1 & 0 \\
0 & 0 & 1 & 0 & 0 & 1 & 0 & 0 \\
0 & 0 & 0 & 1 & 0 & 0 & 0 & 1 
\end{array}}
\right].
\end{equation}

As in \cite{Xu2004}, by modifying the constraints $\bm{g}(\bm{u}^{\prime})$, it is possible to consider zero-injection buses, eliminate critical measurements [see the box \emph{Measurement classification}], and retain observability under a set of contingencies and single PMU losses \cite{Emami2008, Emami2010}. Interestingly, a set of 1,291 PMUs resulted from solving (\ref{eq.13a}) for a large utility system with 2,285 buses. Note that 1,291 > $n_{bu}/2$, which is, in principle, the worst-case scenario \cite{Baldwin1993}. This result provides evidence that the minimum number of PMUs required to attain system observability might be considerably augmented by considering practical factors. This is clearly the case of devices with limited measurement channels. A set of devices with more measurement channels tends to yield a larger spanning measurement tree. For other works that consider the case of limited measurement channels, see \cite{Korkali2009b, Gomez2014}.

\begin{tcolorbox}[colback=black!5!white,
colframe=white!20!black,
title=\textsc{Measurement classification},
center, 
valign = center, 
halign = justify,
before skip = 0.1cm, 
after skip = 0.1cm,
center title, 
width = 1.0\linewidth,
left = 0.1cm,
right = 0.1cm,
before upper = {\parindent1em},
floatplacement = t, float]

In static state estimation, a measurement is classified as either \textbf{critical} or \textbf{redundant}. A well-designed measurement system should not contain critical measurements. See, e.g., \cite{Exposito1998} for more details.

\vspace{.1cm}
A measurement is \textbf{critical} if, after removing it, (\ref{eq.sysmeasmodel}) becomes unobservable. Errors in critical measurements cannot be detected by bad data analysis algorithms embedded in the state estimator.

\vspace{.1cm}
A measurement is \textbf{redundant} if, after removing it, (\ref{eq.sysmeasmodel}) remains observable. Errors in redundant measurements can always be detected by bad data analysis.
\end{tcolorbox}

\subsubsection{Measurement redundancy for static state estimation in transmission grids}\label{subsec.RedundancyT}$\,$

\vspace{.1cm}
To this point, the review is centered around the question of the minimum set of PMUs that makes (\ref{eq.sysmeasmodel}) observable. For state estimation, however, the observability of the measurement model is not enough. One key task performed by a state estimator is to detect, identify, and correct measurement errors. This task is referred to as bad data analysis \cite{Handschin1975}, and it is improved by measurement redundancy. The elimination of critical measurements is of particular interest. This is because the state estimator is unable to detect if a critical measurement is a bad measurement.

A PMU placement strategy that eliminates critical measurements in the existing RTU measurement system is developed in \cite{Chen2005} and extended in \cite{Chen2006}. The work in \cite{Chen2005, Chen2006} develops and expands upon the applicability of the IP method. It starts with a clever observation that from a topological standpoint, the system observability is independent of the numerical value of the parameters, $\{y_{k0},y_{k\ell}\}$, and the algebraic state vector, $\bm{x}$. This allows for series admittances to be set equal to $j1.0$ per unit and voltage magnitudes to be set equal to $1.0$ per unit. Shunt admittances are neglected. We stress that (topological) observability analysis does not depend on the actual state of the system or the branch parameters, making it possible to use these simplifications without loss of generality \cite{Abur2004}. By plugging values into:

\vspace{-.3cm}
\begin{equation}\label{eq.9b}
i_{k\ell} = y_{k\ell}\left(|v_{k}|e^{j\theta_{k}} - |v_{\ell}|e^{j\theta_{\ell}}\right) = -j1.0\left(e^{j\theta_{k}} - e^{j\theta_{\ell}}\right),
\end{equation}

\vspace{-.1cm}
\noindent
and after some algebraic manipulations, one obtains a linear regression model between currents and voltage phase angles:

\vspace{-.3cm}
\begin{equation}\label{eq.10b}
\text{Re}(i_{k\ell})\approx \theta_{k}-\theta_{\ell},
\end{equation}

\vspace{-.1cm}
\noindent
that is valid for sufficiently small $\theta_{k}$ and $\theta_{\ell}$, such that:

\vspace{-.6cm}
\begin{equation}\label{eq.26z}
\left[{\def\arraystretch{0.7}
\begin{array}{l}
\vdots \\
\theta_{k} \\
\text{Re}(i_{k\ell}) \\
\vdots
\end{array}}
\right] = 
\left[{\arraycolsep=2.0pt\def\arraystretch{0.8}
\begin{array}{rrrr}
\ddots &   &    &         \\
\cdots & 1 &  0 & \cdots  \\
\cdots & 1 & -1 & \cdots  \\
       &   &    & \ddots
\end{array}}
\right]
\left[{\def\arraystretch{0.8}
\begin{array}{c}
\vdots \\
\theta_{k} \\
\theta_{\ell} \\
\vdots
\end{array}}
\right] = \bm{H}_{pmu} \cdot \bm{\theta}.
\end{equation}

\vspace{-.2cm}
Note that the algebraic manipulations in (\ref{eq.9b})--(\ref{eq.26z}) are necessary to merge the measurement model of the RTU measurement system, which is nonlinear, with the measurement model of the PMU measurement system, which is linear. Next, instead of (\ref{eq.sysmeasmodel}), consider the measurement model:

\vspace{-.3cm}
\begin{equation}\label{eq.11b}
\bm{z}_{P} = \bm{H}_{P\theta}\bm{\theta} + \bm{e},
\end{equation}

\vspace{-.1cm}
\noindent
where $\bm{z}_{P}$ contains measurements of real power flows and injections, and $\bm{H}_{P\theta}$ is the $P\theta$ sub-matrix of the Jacobian matrix obtained from the DC power flow equations. See also \cite{Gou2000, Donmez2011}. By augmenting (\ref{eq.11b}) with (\ref{eq.10b}):

\vspace{-.2cm}
\begin{equation}\label{eq.13b}
\text{Measurement matrix} = \left[
\begin{array}{l}
\bm{H}_{P\theta} \\
\bm{H}_{pmu}
\end{array}
\right],
\end{equation}

\vspace{-.1cm}
\noindent
and it is possible to consider mixed measurement systems of RTUs and PMUs together. The measurement matrix in (\ref{eq.13b}) is used along with the IP method to place PMUs and eliminate critical measurements simultaneously. See \cite{Chen2006} for more details and \cite{London2008, Chen2008a} for an insightful discussion. The computational efficiency of the method in \cite{Chen2006} can be improved by resorting to the lower-upper decomposition of the measurement matrix along with sparsity techniques \cite{Donmez2011}.

In addition to eliminating critical measurements, one must consider leverage measurements [see the box \emph{Leverage measurement}]. This is because leverage measurements have an undue effect on most power system state estimators, with a few exceptions \cite{Mili1991, Mili1996, mili1990b}. Note that most energy management systems available commercially employ a WLS state estimator, which is vulnerable to leverage measurements. Indeed, the vulnerability of the WLS estimator to bad data has been known since the seminal work of Schweppe and Handschin \cite{Schweppe1974, Handschin1975}. Moreover, as pinpointed in \cite{Mili1991}, locations susceptible to create leverage measurements ``have to be provided with enough measurements in order to increase their local redundancy. Indeed, they tend to be isolated in the factor space and weakly coupled with the surrounding measurements.'' Thus, a cluster of measurements around these locations is recommended to bound their influence in the estimation process; otherwise, the algebraic states (voltage phasors) close to these locations are estimated with large variances. Unfortunately, though the formulation in \cite{Chen2005, Chen2006} eliminates critical measurements, it does not consider leverage measurements. The problem of measurement redundancy is only partially solved in \cite{Chen2005, Chen2006}. The consideration of leverage measurements is a gap in the literature and an opportunity for future development. This is further discussed in Section \ref{section.discussion}.

\begin{tcolorbox}[colback=black!5!white,
colframe=white!20!black,
title=\textsc{Leverage measurement},
center, 
valign = center, 
halign = justify,
before skip = 0.1cm, 
after skip = 0.1cm,
center title, 
width = 1.0\linewidth,
left = 0.1cm,
right = 0.1cm,
before upper = {\parindent1em},
floatplacement = t, float]

\noindent
Recall the measurement model (\ref{eq.sysmeasmodel}), which is given by $\bm{z} = \bm{H}\bm{x} + \bm{e}$. Let $z_{i}$ be the $i$-th element of $\bm{z}$, and $\bm{h}_{i}\tran$ be the $i$-th row of $\bm{H}$. All the row vectors $\bm{h}_{i}\tran$, $i=\{1,...,m\}$, lie in the so-called factor space of regression. If there is an outlier in $\bm{h}_{i}\tran$, then the corresponding measurement $z_{i}$ will have an undue influence on the state estimate $\widehat{\bm{x}}$. See (\ref{eq.24c}). In this case, $z_{i}$ is a \emph{leverage measurement}. For more details, the reader is referred to \cite{Mili1991, Abur2004}. Note that this definition of a leverage measurement is based on the notion of factor space, and it is anchored on statistics theory. The notion of leverage measurements was introduced to the power system community by Mili, Rouseeuw, and colleagues \cite{Mili1991}. So-called by statisticians as influential data points, Rouseeuw and colleagues \cite{Rousseeuw1990, Rousseeuw2005} coined the term \emph{leverage point}. Note, however, that the numerical effect of leverage measurements on power system state estimation was noticed earlier by Monticelli \cite{Monticelli1999}. The following conditions are known to create leverage measurements in power systems:

\vspace{-.2cm}
\begin{itemize}
\item An injection measurement placed at a bus that is incident to a large number of branches
\item An injection measurement placed at a bus that is incident to branches of very different impedance values
\item Flow measurements along branches whose impedances are very different from those of the other branches in the system
\item Using a very large weight for a specific measurement.
\end{itemize}
\end{tcolorbox}

\vspace{.2cm}
\noindent
\emph{Consideration of topology changes and topology errors:}

\vspace{.1cm}

In addition to critical measurements, topology changes are another factor that might render a measurement system with minimally placed PMUs ineffective. Note that topology changes occur every so often, and they are driven by, e.g., scheduled maintenance of equipment, seasonal trends in electric power consumption, and unforeseen equipment/systems outages. 

Additionally, the measurement matrix, $\bm{H}$, needs to be reevaluated after any topology change. This is accomplished by topology processing algorithms embedded into the static state estimators. These algorithms are responsible for building the network connectivity model---also referred to as a bus-branch model---based on the status of switchgear in the field; however, it is not uncommon for the equipment statuses to be reported incorrectly, e.g., because of communications issues. This is an issue because topology errors yield biased estimates and might cause the state estimator to diverge.

Therefore, a reliable measurement system should guarantee observability in case of topology changes and in case of erroneous switchgear status \cite{Clements1988}. By relying on the original ideas in \cite{Baldwin1993}, \cite{Denegri2002} addresses N-1 contingencies, and \cite{Milosevic2003} addresses single branch outages. In general, to ensure topological observability in case of contingencies, additional constraints can be imposed on the IP formulation as follows:

\vspace{-.3cm}
\begin{equation}
g_{k}^{c}(\bm{u})=\sum_{\ell=1}^{n_{bu}}A_{k\ell}^{c} \cdot u_{\ell}\ge 1,
\end{equation}

\vspace{-.1cm}
\noindent
where the constraint $g_{k}^{c}$ enforces the topological observability of bus $k$ in case of a contingency $c$, e.g., the loss of a branch or a PMU; $A_{k\ell}^{c}$ denotes the $k\ell$-th element of the bus-to-bus connectivity matrix of the system under a contingency; and $u_{\ell}$ is defined as in (\ref{eq.4b}). Also, some work has addressed the case of single outages by enforcing all buses to be observed at least twice by PMUs---to do so, one simply sets the right-hand side of (\ref{eq.4a}) to be $\ge 2$ instead of $\ge 1$. 

The work in \cite{Rakpenthai2007} builds on \cite{Madtharad2003} and addresses single measurement losses and single branch outages. These strategies remedy specific cases of topology changes but do not address the case of erroneous switchgear status. Conversely, \cite{Chen2008b} expands on (\ref{eq.9b})--(\ref{eq.13b}) and develops a PMU placement strategy that ensures that any single branch topology error is detectable. As a by-product of the developed strategy, it is guaranteed that the system will remain observable in case of any single branch contingency. The approach in \cite{Chen2008b} provides a systematic way to address the problem of topology changes and is overall superior to the others. Note that the approach in \cite{Chen2008b} does not account for leverage measurements either. 

Other ideas with limited impact appear in the literature. An example is considering controlled islanding. Provided the system has a sufficient degree of observability, including controlled islanding \cite{Huang2014a} in the measurement placement problem is unnecessary.

\begin{tcolorbox}[colback=black!5!white,
colframe=white!20!black,
title=\textsc{Residual sensitivity analysis},
center, 
valign = center, 
halign = justify,
before skip = 0.1cm, 
after skip = 0.1cm,
center title, 
width = 1.0\linewidth,
left = 0.1cm,
right = 0.1cm,
before upper = {\parindent1em},
floatplacement = b, float]

Define $\widehat{\bm{z}}:=\bm{H}\widehat{\bm{x}}$. Then, using (\ref{eq.24c}):

\vspace{-.3cm}
\begin{equation}\label{eq.27d}
\widehat{\bm{z}} = \bm{H}\left(\bm{H}\tran \bm{R}^{-1}\bm{H}\right)^{-1}\bm{H}\tran \bm{R}^{-1}\bm{z} = \bm{S}\bm{z},
\end{equation}

\vspace{-.1cm}
\noindent
where $\bm{S}$ is referred to as the \emph{hat matrix}. Now, define the residual vector $\bm{r} := \bm{z} - \widehat{\bm{z}}$. Thus:

\vspace{-.5cm}
\begin{align}
\bm{r} &= \bm{z} - \bm{S}\bm{z} = \left(\bm{I}-\bm{S}\right)\bm{z} = \bm{W}\bm{z} = \bm{W}\left(\bm{H}\bm{x}+\bm{e}\right) \nonumber \\
&= \bm{W}\bm{H}\bm{x} + \bm{W}\bm{e} = \left(\bm{I}-\bm{S}\right)\bm{H}\bm{x} + \bm{W}\bm{e} \nonumber \\
&= \left(\bm{H}\bm{x}-\bm{H}\left(\bm{H}\tran \bm{R}^{-1}\bm{H}\right)^{-1}\bm{H}\tran \bm{R}^{-1}\bm{H}\bm{x}\right)+\bm{W}\bm{e} \nonumber \\
&= \bm{W}\bm{e},
\end{align}

\vspace{-.2cm}
\noindent
where $\bm{W}$ is referred to as the \emph{residual sensitivity matrix}. Now, under the assumption that the measurement errors are Gaussian:

\vspace{-.5cm}
\begin{align}
\mathcal{E}(\bm{r}) &= \mathcal{E}(\bm{W}\bm{e}) = \bm{W}\mathcal{E}(\bm{e}) = \bm{0}, \\
\mathcal{E}(\bm{r}\bm{r}\tran ) &= \mathcal{E}(\bm{W}\bm{e}\bm{e}\tran \bm{W}\tran ) = \bm{W}\mathcal{E}(\bm{e}\bm{e}\tran )\bm{W}\tran = \bm{W}\bm{R}\bm{W}\tran . \label{eq.30c}
\end{align}

\vspace{-.5cm}
It can be verified that $\left(\bm{R}\bm{W}\tran \right)\tran =\bm{R}\bm{W}\tran $. By plugging this into (\ref{eq.30c}) and using the fact that $\bm{R}$ is a diagonal matrix:

\vspace{-.6cm}
\begin{align}\label{eq.31d}
\mathcal{E}(\bm{r}\bm{r}\tran ) &= \bm{W}\bm{W}\bm{R} = \bm{W}\bm{R},
\end{align}

\vspace{-.2cm}
\noindent
where $\bm{W}\bm{W}=\bm{W}$ because $\bm{W}$ is an idempotent matrix; hence, the residual sensitivity is related to the error covariance matrix, $\bm{R}$.
\end{tcolorbox}

\subsubsection{State estimation accuracy under ideal conditions}\label{subsec.AccuracyT}$\,$

\vspace{.1cm}
There is prolific work on the minimum placement of PMUs for system observability. Some also consider the legacy measurement system based on RTUs. For example, \cite{Chen2005, Chen2006} develops a systematic approach to eliminate critical measurements in the legacy measurement system by strategically allocating PMUs. Following \cite{Chen2005, Chen2006}, the work in \cite{Li2011} also supposes that the system is observable through the legacy measurement system, but it brings in a new perspective---it focuses on placing PMUs to reduce the variance of estimated quantities [see the box \emph{Residual sensitivity analysis}]. By reducing the variance of estimated quantities, one improves the accuracy of the state estimation process. 

It is cumbersome to consider variance reduction in formulations using the notion of topological observability. This is the case, for example, of the previously described IP method and its extensions. An error covariance matrix exogenous to the original formulation would have to be built and assessed iteratively. Conversely, the notion of numerical observability lends itself better by intrinsically considering the covariance matrix within the formulation. In this context, the idea of using the covariance matrix to place measurements can be traced back to the seminal work of Fred C. Schweppe and colleagues; see, e.g., \cite{Schweppe1970} (page 124, \emph{Discussion of Theory}), or \cite{Schweppe1974} (page 980, \emph{Meter Configuration}). These works use the numerical observability-based problem formulation. 

From the standpoint of an optimal experimental design, a reasonable goal during the design phase of an experiment is to minimize $\bm{G}^{-1}$ in some way. Note that $\bm{G}^{-1}$ has a direct impact on the state estimate $\widehat{\bm{x}}$; see (\ref{eq.24c}). There are many different ways in which $\bm{G}^{-1}$ might be made minimal. Three optimality criteria considered in \cite{Li2011} are:

\vspace{-.3cm}
\begin{itemize}
\item \emph{A-optimality}: minimize the trace of $\bm{G}^{-1}$.
\item \emph{D-optimality}: minimize the determinant of $\bm{G}^{-1}$.
\item \emph{E-optimality}: minimize the maximum eigenvalue of $\bm{G}^{-1}$.
\end{itemize}

\vspace{-.3cm}
The work in \cite{Li2011} is extended in \cite{Li2013, Shi2020}, wherein the mutual information is the adopted criterion. Both works use a greedy algorithm to solve the problem numerically. The idea of placing PMUs to improve the state estimation performance from an optimal experimental design standpoint is also pursued in \cite{Kekatos2012, Li2014, Korres2015}, wherein the problem is solved after relaxing it to a convex semidefinite program. In particular, \cite{Li2014} formulates a multicriteria framework that concurrently seeks to:

\vspace{-.3cm}
\begin{itemize}
\item Increase measurement redundancy
\item Eliminate critical measurements
\item Maintain observability in case of single branch outages
\item Attain \emph{E-optimality}
\item Improve the convergence of the Gauss-Newton algorithm.
\end{itemize}

\vspace{-.3cm}
The latter is noteworthy because \cite{Li2014} is the first attempt to consider the numerical convergence of the state estimation solution process in the PMU placement problem. Note, however, that the iteratively reweighted least-squares algorithm \cite{Pires1999} is preferred over the standard Gauss-Newton algorithm because of superior numerical stability. See also \cite{Yang2015}, wherein the optimal design considers the fact that PMU measurements of phase angles (for both voltages and currents) are not perfectly synchronized; and, \cite{Sun2019}, wherein the robust LAV estimator is considered in addition to the classic WLS estimator. The works in \cite{Li2011, Li2013, Shi2020, Kekatos2012, Li2014, Korres2015, Yang2015, Sun2019} also offer interesting theoretical developments, particularly in the areas of signal processing and optimization. 

In power systems, however, the standard deviations of the noise of the metering devices and their associated communications channels, including the PMUs, are estimated with large uncertainties \cite{Mili1985, Jaen2018}. This fact adds to the level of uncertainty of any approach that, directly or indirectly, uses $\bm{G}^{-1}$. Moreover, to each state $\bm{x}$, there is a corresponding matrix $\bm{G}^{-1}$. In other words, $\bm{G}^{-1}$ varies depending on the system operating condition. The use of a single matrix $\bm{G}^{-1}$ is, therefore, discouraged because it yields a plan that focuses on a single operating scenario. This drawback can nonetheless be circumvented by using multiple operating conditions sampled, e.g., through Monte Carlo simulations.

\begin{figure*}[!ht]
\centering
\begin{tikzpicture}[
 bigcircle/.style={ 
    text width=1.6cm, 
    align=center, 
    line width=2mm, 
    draw, 
    circle, 
    font=\sffamily\scriptsize 
  },
 desc/.style 2 args={ 
  text width=4.0cm, 
  font=\sffamily\tiny\RaggedRight, 
  label={[#1,yshift=-1.5ex,font=\sffamily\scriptsize]above:#2} 
  },
 node distance=10mm and 2mm 
]

\node [bigcircle] (circ1) {Identify main cost components};
\node [desc={black}{Cost of},below=of circ1] (list1) {
\begin{itemize}
\setlength\itemsep{0pt} 
\item Measurement devices
\item Instrument transformers
\item Communications
\item Data storage
\end{itemize}
};

\node [bigcircle,red,right=of list1] (circ2) {Define candidate locations};
\node [desc={red}{Analyze planning model to eliminate},above=of circ2] (list2) {
\begin{itemize}
\setlength\itemsep{0pt}
\item Virtual buses
\item Radial buses
\item Zero-injection buses
\end{itemize}
};

\node [bigcircle,blue!60!red,right=of list2] (circ3) {Solve basic problem};
\node [desc={blue!60!red}{For minimal observability considering},below=of circ3] (list3) {
\begin{itemize}
\setlength\itemsep{0pt}
\item Main cost components
\item Preselection of candidate buses
\item Zero-injection buses
\end{itemize}
};

\node [bigcircle,gray,right=of list3] (circ4) {Define strategy};
\node [desc={gray}{Based on},above=of circ4] (list4) {
\begin{itemize}
\setlength\itemsep{0pt}
\item Minimal observability solution
\item Budget
\item Project timeline
\end{itemize}
};

\node [bigcircle,blue,right=of list4] (circ5) {Solve\\multistage problem};
\node [desc={blue}{Considering},below=of circ5] (h) {
\begin{itemize}
\setlength\itemsep{0pt}
\item Main cost components
\item Preselection of candidate buses
\item Zero-injection buses
\item Existing infrastructure
\item Critical measurements
\item Leverage measurements
\item Loss of measurement devices
\item Loss of communications link
\item Branch outages
\item Critical buses priority
\item Estimation accuracy
\end{itemize}
};

\draw [dashed,black!80] (circ1) -- (circ2) -- (circ3) -- (circ4) -- (circ5);
\end{tikzpicture} 
\vspace{-.2cm}
\caption{Steps and summary of the most important factors to consider in the measurement placement problem. The focus is on only static state estimation in transmission grids.}
\label{fig.summary.SSE}
\end{figure*}
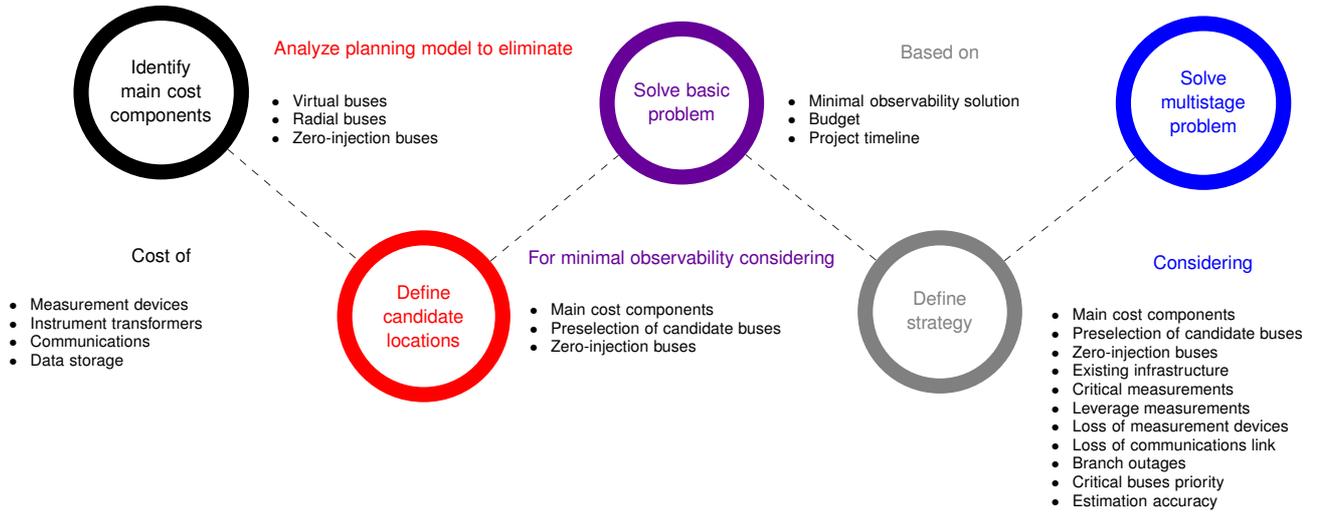

\vspace{.2cm}
\noindent
\emph{Consideration of additional costs:}

\vspace{.1cm}
The discussion until now focuses on the cost of placing PMUs subject to constraints, which are specifically tailored to guarantee a predefined goal, e.g., system observability. But a study conducted by the U.S. Department of Energy reveals that the cost of PMUs represents approximately 5\% of the total investment \cite{ARRA2014}. The communications infrastructure is unquestionably an important layer between the measurement devices and the end-user application \cite{Liu2000}. This aspect is overlooked in most of the previously discussed work. A few exceptions include \cite{Nuqui2005, Rakpenthai2007}, in which this aspect is superficially touched upon. The work in \cite{Shahraeini2012} accounts for an attempt to expand on that front by co-optimizing the cost of PMUs and the cost of building the communications infrastructure. The cost of the communications infrastructure is explicitly formulated as follows:

\vspace{-.3cm}
\begin{equation}\label{eq.27c}
\begin{aligned}
\min_{} \quad & \sum_{k=1}^{n_{link}}c_{link,k} \cdot \lambda_{k},
\end{aligned}
\end{equation}

\vspace{-.2cm}
\noindent
where $n_{link}$ is the number of communications links necessary to form a connected graph with the PMUs; $c_{link,k}$ is the cost per unit of length of the communications link $k$; and $\lambda_{k}$ is the length of the communications link $k$. The IP method is augmented with (\ref{eq.27c}), and the solution is obtained through a multi-objective genetic algorithm. Further costs are considered in \cite{Aminifar2013, Rather2015, Mohammadi2016, Singh2017}. In particular, \cite{Rather2015} reports on the experience obtained from the expansion of the wide-area measurement system of the Danish transmission grid. Although most extensions of the IP method overlook the dissimilarity in the cost of installing PMUs on different buses and focus on the set of constraints in the pursuit of specific enhancements, \cite{Rather2015} focuses on the overall cost and reveals a set of important hidden factors that must be included in the optimization model. The discussions related to instrument transformers are particularly noteworthy. The cost of a set of three potential transformers (PTs) plus three current transformers (CTs)---one PT and CT per phase in a three-phase system---is approximately six times the cost of a PMU. This does not include the cost of the structural foundation that is necessary to install the instrument transformers in the substation. Nonetheless, multichannel PMUs require three CTs per current measurement channel. Also, there are additional hidden costs. For instance, it might be necessary to curtail generation or load for several hours to install the measurement system, from which shutdown costs will incur. It is interesting that though critical facilities such as large generation plants are taken as preferred locations to install PMUs, they might incur the highest shutdown costs. Also, the installation of instrument transformers involves a significant amount of man-hours for engineering, cabling, etc. Fortunately, it is straightforward to include these costs in the formulation of the IP method. Reference \cite{Rather2015} also provides a cost baseline that might be useful for initial projections. For costs associated with substation infrastructure, see \cite{Pal2017b, Almasabi2018}. For an informative discussion on the communications infrastructure, see \cite{Appasani2018}. See also \cite{Sarailoo2018, Zhu2019}.

\vspace{.2cm}
\noindent
\emph{Preselection of the set of candidate buses:}

\vspace{.1cm}
In practice, before attempting to solve a heuristic or an optimization model, it is necessary to build a set of candidate locations to place measurements. In this process, some buses might be selected for mandatory PMU installations, whereas others can be eliminated, thereby reducing the effort to search for an optimal solution. The latter is very important because the problem of finding the minimum PMU placement for system observability is NP-complete \cite{Brueni2005}. In the models of electric power grids, it is common to encounter buses that do not exist physically or are not in practical locations to install PMUs. The elimination of these buses, though it does not affect the final solution, significantly reduces the effort to find it \cite{Zhou2008}. Radial buses are another important consideration. For example, in the network in Fig. \ref{fig.7bus}, buses $\{1,5\}$ are radial buses. Consider Bus 1. To make Bus 1 observable, there are only two candidate buses to pick from: $\{1,2\}$. Installing a PMU on Bus 1 leads to buses $\{1,2\}$ being observable, whereas installing a PMU on Bus 2 leads to buses $\{1,2,3,6,7\}$ being observable. The second option is better. This observation can always be leveraged to eliminate radial buses from the candidate set. Another important observation relates to zero-injection buses. As discussed in previous sections, the consideration of zero-injection buses leads to a reduction in the set of constraints; thus, in general, it is recommended to not include zero-injection buses in the set of candidate buses.

Before moving on to the next section, we present a summary of important factors. See Fig. \ref{fig.summary.SSE}. These factors must be addressed when the focus of the measurement placement problem is on static state estimation.

{\color{black}
\vspace{-.2cm}
\subsection{Dynamic state estimation in transmission grids}\label{sec.DynamicStateEstimation}
}

Dynamic state estimation has gained momentum in the power system community. Static and dynamic state estimation are significantly different in many aspects, such as modeling and assumptions, the range of applications, and the requirements imposed on the measurement systems and communications networks. See, e.g., \cite{Netto2018, Netto2019, Zhao2019, Zhao2020}. Here, the important difference between the two is in the measurement model (\ref{eq.sysmeasmodel}), specifically, in the vector $\bm{x}$. For \emph{static state estimation}, $\bm{x}$ denotes the algebraic state vector, which contains the voltage magnitude in all buses and the voltage phase angle in all but the reference bus. As for \emph{dynamic state estimation}, $\bm{x}$ denotes the dynamic state vector or simply the ``state vector'' (the latter is more common outside the power system community). In this case, $\bm{x}$ contains the variables associated with generators and their controllers, for example, the rotor speed of synchronous generators and the pitch angle of wind turbines. Because $\bm{x}$ is different, $\bm{H}$ is different. More importantly, for dynamic state estimation, $\bm{H}$ is time-varying. The notion of observability is thus more involved. In the case of dynamic state estimation, as articulated in \cite{Rouhani2017}, \emph{``higher (lower) values of the smallest singular value of the observability matrix indicates stronger (weaker) observability for a given measurement set. Since observability is a local property, the smallest singular value of the observability matrix will change along the trajectory of $\bm{x}$.''} In contrast, the notion of observability in static state estimation is binary---that is, either the system is observable or not observable---and not time-varying; therefore, the problem of observability for dynamic state estimation is more challenging, as is the problem of measurement placement that seeks to attain observability.

The work in \cite{Zhang2010} is the first to formulate a PMU placement problem that considers dynamic state estimation. The classical synchronous generator model \cite{Machowski2020} is adopted, and the strategy is as follows. In the first stage, the IP method is executed to obtain the minimum set of PMUs that yields (\ref{eq.sysmeasmodel}) as topologically observable for static state estimation; this typically leads to multiple solutions. Then, in the second stage, the solutions are ranked using a criterion that relates to dynamic state estimation. The criterion is based on the asymptotic error covariance matrix of the Kalman filter, given by:

\vspace{-.3cm}
\begin{equation}
\lim_{k \to \infty} \mathcal{E}\left[(\bm{x}-\widehat{\bm{x}})(\bm{x}-\widehat{\bm{x}})\tran \right],
\end{equation}

\vspace{-.1cm}
\noindent
where $\bm{x}$ and $\widehat{\bm{x}}$ denote, respectively, the true and estimated \emph{dynamic} state vector. Note that although the Gauss-Newton or the iteratively reweighted least-squares algorithms are used in the solution process of the static state estimation, dynamic state estimation relies on Kalman filtering. The work in \cite{Tai2013} proposes the use of lower and upper bounds instead of the asymptotic error covariance matrix. The work in \cite{Qi2015} does not consider static state estimation and focuses exclusively on dynamic state estimation. It formulates an optimization model to maximize the determinant of the empirical observability Gramian, which is calculated for a set of operating points to quantify the degree of observability of a given PMU placement. See also \cite{Sun2011}.

Dynamic state estimation will play a critical role in power systems control and protection \cite{Liu2021}. Despite the availability of preliminary results, the problem of measurement placement for dynamic state estimation remains open. Accordingly, measurement placement for dynamic state estimation is another opportunity for future development. This is further discussed in Section \ref{section.discussion}.

\vspace{.2cm}
\subsection{Stability assessment and control in transmission grids}\label{sec.StabilityAssessmentAndControlT}

Real-time stability \cite{Kundur2004, Hatziargyriou2020} assessment and control represent another major application of PMUs. This is discussed next.

\subsubsection{Voltage control in transmission grids }\label{subsec.VoltageT}$\,$

\vspace{.1cm}
The work in \cite{Mili1990} studies the effect of PMU location on the secondary voltage control of transmission grids; it argues that the minimum set of PMUs that makes (\ref{eq.sysmeasmodel}) observable is sufficient for systematically identifying pilot points. Note that, following \cite{Mili1990}, a \emph{``pilot point is a voltage at a load bus which is measured in real-time and used for control action.''} The pilot point must be representative of all the voltages within the region where it is located. The adoption of the notion of pilot points, however, might be obsolete in the context of the ongoing modernization of electric power grids. This is because of the sustained growth in the number of flexible AC transmission system devices deployed worldwide. Also, an investigation of the effect of incremental PMU placement on decision tree-based online voltage security monitoring \cite{Nuqui2001} reveals an overall improvement in voltage security misclassification rates.

The work in \cite{Kumar2016} combines the multistage ILP method with a measure of the system dynamic performance. The idea is to perform time-domain simulations and rank the system buses according to their dynamic vulnerability. The buses are divided into \emph{generator buses} and \emph{load buses}. The vulnerability of the generator buses is calculated using the individual machine energy functions \cite{Michel1983}. The vulnerability of the load buses is calculated using the concept of proximity to voltage collapse \cite{Vu1999}. Based on the obtained bus ranking, the most vulnerable buses are made observable at the earlier stages of the multistage placement. A similar idea is pursued in \cite{Rashidi2016}, wherein buses are ranked based on their correspondence to the largest Lyapunov exponent of the network \cite{Dasgupta2013}. The idea is that buses with a strong correspondence to the largest Lyapunov exponent are more significant from the system stability standpoint, and they should be made observable with higher priority. See also \cite{Castillo2020}. The drawback of these approaches lies in their dependence on an explicit energy function, which is challenging to obtain.

\subsubsection{Rotor angle stability assessment in transmission grids}\label{subsec.RotorAngle}$\,$

\vspace{.1cm}
The work in \cite{Palmer1996} focuses on (small-signal) rotor angle stability. Consider the similarity transformation:

\vspace{-.3cm}
\begin{equation}\label{eq.similarity-transformation}
\bm{w} = \bm{U}^{-1}\bm{x},
\end{equation}

\vspace{-.2cm}
\noindent
where $\bm{w}$ denotes the vector of the modal variables, and $\bm{U}$ is the matrix containing the right eigenvectors of the linearized system model. By plugging (\ref{eq.similarity-transformation}) into (\ref{eq.sysmeasmodel}), one obtains: 

\vspace{-.3cm}
\begin{equation}
\bm{z} = \bm{H}\bm{U}\bm{w} + \bm{e},
\end{equation}

\vspace{-.2cm}
\noindent
a formal relationship between the measurements and modal variables. The key idea in \cite{Palmer1996} is to use the matrix product $\bm{H}\bm{U}$ to find a set of measurements from which all inter-area modes are observable. This method has two important drawbacks: 

\vspace{-.3cm}
\begin{itemize}
\item Its performance on highly meshed networks might be poor. Note that electric power transmission grids are typically highly meshed.
\item It relies on information obtained from the eigendecomposition of the linearized system model. This information can vary significantly under different operating scenarios.
\end{itemize}

\vspace{-.3cm}
The work in \cite{El-Shal1990} develops a simplistic approach to place PMUs on tie-lies with the objective of monitoring the internal voltage of synchronous generators. It focuses on the (transient) rotor angle stability of reduced two-area power systems, and because of that, it has limited applicability. A more elaborate strategy is available in \cite{Kamwa2002}, which starts by considering the set of all network buses, $\mathbb{B}$, $\text{card}(\mathbb{B})=n_{bu}$; a set of candidate buses, $\mathbb{C}$, $\text{card}(\mathbb{C})=n_{c}$; and a set of credible disturbances, $\mathbb{D}$, $\text{card}(\mathbb{D})=n_{d}$. Then, for a given disturbance $d\in\mathbb{D}$, define the measurement matrix:

\vspace{-.3cm}
\begin{equation}
\bm{Z}_{\mathbb{C}}^{(d)} = \left[ \bm{z}_{1}(t) \;\; \bm{z}_{2}(t) \;\; ... \;\; \bm{z}_{n_{c}}(t) \right],
\end{equation}

\vspace{-.1cm}
\noindent
where $\bm{z}_{k}(t)$ is a vector containing $m$ pseudo-measurements of the bus variable associated with the $k$-th element of $\mathbb{C}$. These pseudo-measurements are generated through the numerical simulation of each disturbance $d\in\mathbb{D}$. The strategy is to select the set of PMUs, $\mathbb{S} \subseteq \mathbb{C} \subseteq \mathbb{B}$, that maximizes the information content of $\bm{Z}_{\mathbb{C}}^{(d)}$, denoted by $\mathcal{I}\left(\bm{Z}_{\mathbb{C}}^{(d)}\right)$, which is quantified by some norm (typically the $\ell^2$-norm) of:

\vspace{-.3cm}
\begin{itemize}
\item The entropy matrix, $\bm{E}$---that is, $\mathcal{I}\left(\bm{Z}_{\mathbb{C}}^{(d)}\right)=||\bm{E}||$; or
\item The coherency matrix, $\bm{C}$---that is, $\mathcal{I}\left(\bm{Z}_{\mathbb{C}}^{(d)}\right)=||\bm{C}||$.
\end{itemize}

\vspace{-.3cm}
The choice in \cite{Kamwa2002} is the Gramian norm, which is defined as:

\vspace{-.5cm}
\begin{equation}
\left|\left|\mathcal{I}\left(\bm{Z}_{\mathbb{C}}^{(d)}\right)\right|\right| = \text{det}\left(\left[\mathcal{I}\left(\bm{Z}_{\mathbb{C}}^{(d)}\right)\right]\tran \left[\mathcal{I}\left(\bm{Z}_{\mathbb{C}}^{(d)}\right)\right]\right)^{1/n_{c}}.
\end{equation}

\vspace{-.2cm}
The $k\ell$-th element of $\bm{E}$ is given by:

\vspace{-.2cm}
\begin{equation}
E_{k\ell} = {E}_{\ell k} = \int_{0}^{\infty} \log \left( \frac{\bm{z}_{k\ell}(\omega)}{\sqrt{\bm{z}_{kk}(\omega)\bm{z}_{\ell\ell}(\omega)}} \right)d\omega,
\end{equation}

\vspace{-.1cm}
\noindent
where $\bm{z}_{k\ell}(\omega)$ is the power spectral density for $k=\ell$, and the cross spectral density otherwise. The $k\ell$-th element of $\bm{C}$ is given by:

\vspace{-.4cm}
\begin{equation}
C_{k\ell} = {C}_{\ell k} = \sqrt{\frac{1}{T}\left(\int_{1}^{T}\left[\theta_{k}(t)-\theta_{\ell}(t)\right]^{2} + \left[f_{k}(t)-f_{\ell}(t)\right]^{2}dt\right)},
\end{equation}

\vspace{-.2cm}
\noindent
where $T$ denotes the sampling period, and $f_{k}$ denotes the frequency at bus $k$. The optimization model is defined as:

\vspace{-.2cm}
\begin{equation}
\begin{aligned}
\min_{} \quad & \mathcal{I}\left(\bm{Z}_{\mathbb{C}}^{(d)}\right), \\
\textrm{s.t.} \quad & n_{p} \le n_{c},
\end{aligned}
\end{equation}

\vspace{-.1cm}
\noindent
where $n_{p}$ denotes the number of placed PMUs. See also \cite{Lara-Jimenez2017}.

The consideration of other notions of stability---particularly frequency, resonance, and converter-driven stability \cite{Hatziargyriou2020}---is a gap in the literature and an opportunity for future development. This is further discussed in Section \ref{section.discussion}.

\vspace{-.2cm}
\subsection{Topology change detection in transmission grids}\label{sec.TopologyChangeDetectionT}

\vspace{-.2cm}
There have been some attempts to use PMU measurements to detect line outages directly, without requiring state estimation. Most available methods assume that if a line outage occurs, then the voltage phase angles change significantly in response to the change in topology. The work in \cite{Zhao2012} makes the additional assumption that the power injections of the network remain the same within a few seconds after a line outage occurs. Also, it employs the DC power flow model to do offline simulations and collect signatures of the system's voltage phase angle responses to single-line outages. Based on these assumptions, the optimization objective is established as one of maximizing the minimum distance among the voltage phase angle signatures of the outages. The problem is formulated as an IP and solved by using a greedy algorithm. Interestingly, for the IEEE 30-bus system, if 10 PMUs are to be installed to detect line outages, the optimal locations are found to be at buses $\{\underline{1},\underline{5},8,9,14,21,22,\underline{24},26,29\}$. See also \cite{Zhao2014}. Now, based on the exhaustive search in \cite{Chakrabarti2008}, the minimum number of PMUs to make the IEEE 30-bus system observable under normal operating conditions, considering zero-injection buses, is equal to $7$. If single branch outages are considered, then this number increases to $10$. The optimal locations of PMUs obtained in \cite{Chakrabarti2008} are as follows. Considering:

\vspace{-.2cm}
\begin{itemize}
\item Normal operating conditions: $\{\underline{1},2,10,12,15,\{19\text{ or }20\},27\}$
\item Single branch outages $\{2,3,\underline{5},10,12,15,17,19,\underline{24},27\}$.
\end{itemize}

\vspace{-.2cm}
Note that the optimal locations in \cite{Zhao2012} and \cite{Chakrabarti2008} are hardly comparable. This provides a clear illustration that: i) the optimal solution for a particular application might not be the overall best approach, and ii) considering different strategies for optimization separately might lead to conflicting solutions; therefore, such multiple considerations will be necessary to develop a coordinated placement strategy that is cost-effective and applicable to many use cases. In \cite{Kim2018}, a logistic regression-based method is employed to identify the most influential buses for outage detection. See also \cite{Koochi2020}.

\vspace{-.2cm}
\subsection{Fault detection in transmission lines}\label{sec.FaultDetectionT}

\vspace{-.2cm}
The work in \cite{Lien2006} is the first to propose a PMU placement scheme for fault location. The adopted heuristic is simple, and the algorithm has basically two rules: i) place PMUs on the two buses with the largest number of connected branches; ii) such that between two PMU buses there is a bus with no PMU. The solution obtained with this simple scheme is not unique, and the second rule might lead to more PMUs than the minimum needed for fault location; thus, the solution always needs to be refined. More importantly, the number of PMUs required for fault location is much larger than the minimum number of PMUs required for observability in static state estimation. For example, in the IEEE 14-bus system, 3 PMUs are required for observability \cite{Baldwin1993} and 8 PMUs are required for fault location \cite{Lien2006}. See also \cite{Liao2009, Korkali2013, Li2019}.

\vspace{-.2cm}
\subsection{Power quality monitoring in transmission grids}\label{sec.PowerQualityMonitoringT}

\vspace{-.2cm}
The first work to provide insight into how to place measurements for power quality monitoring is \cite{Heydt1989}, which describes a reverse power flow procedure to identify the source of harmonics in electric power grids. Notice that power quality monitoring requires a specific type of meter, referred to as harmonic meter or power quality meter. The procedure relies on a linear relation between the Fourier transforms of bus voltages, $\bm{v}(\omega)$, and bus injection currents, $\bm{i}(\omega)$, as follows:

\vspace{-.1cm}
\begin{equation}
\bm{i}(\omega) = \bm{Y}(\omega)\bm{v}(\omega),
\end{equation}

\vspace{-.1cm}
\noindent
where $\bm{Y}(\omega)$ is the bus admittance matrix. Let $\bm{i}_{o}(\omega)$, $\bm{v}_{o}(\omega)$ denote a vector of observed or measured quantities, and let $\bm{i}_{u}(\omega)$, $\bm{v}_{u}(\omega)$ denote a vector of unobserved quantities, such that:

\vspace{-.2cm}
\begin{equation}
\bm{i}(\omega) =
\left[\begin{array}{c}
\bm{i}_{u}(\omega) \\
\bm{i}_{o}(\omega)
\end{array}\right], \quad
\bm{v}(\omega) =
\left[\begin{array}{c}
\bm{v}_{u}(\omega) \\
\bm{v}_{o}(\omega)
\end{array}\right]
\end{equation}

\vspace{-.1cm}
It follows that:

\vspace{-.3cm}
\begin{equation}
\left[\begin{array}{c}
\bm{i}_{u} \\
\bm{i}_{o}
\end{array}\right] =
\left[\begin{array}{cc}
\bm{Y}_{uu} & \bm{Y}_{uo} \\
\bm{Y}_{ou} & \bm{Y}_{oo}
\end{array}\right]
\left[\begin{array}{c}
\bm{v}_{u} \\
\bm{v}_{o}
\end{array}\right],
\end{equation}

\vspace{-.1cm}
\noindent
where $(\omega)$ is omitted for simplicity of notation. In \cite{Heydt1989}, the unobserved quantities are estimated in the least-squares sense:

\vspace{-.4cm}
\begin{align}
\widehat{\bm{v}}_{u} &= \bm{Y}_{ou}\tran \left(\bm{Y}_{ou}\bm{Y}_{ou}\tran\right)^{-1} \left(\bm{i}_{o} - \bm{Y}_{oo}\bm{v}_{o}\right), \\
\widehat{\bm{i}}_{u} &= \bm{Y}_{uu}\widehat{\bm{v}}_{u} + \bm{Y}_{uo}\bm{v}_{o},
\end{align}

\vspace{-.2cm}
\noindent
and it is suggested to place measurements to reduce the condition number of the matrix $\bm{Y}_{ou}$. The rationale of this strategy is provided, but no guidelines on how to choose the measurement locations are offered. Yet, existing linear algebra-based algorithms can pinpoint the best variables (measurement locations) to choose from to improve the condition number of a given matrix. See, e.g., \cite{VanHuffel1987}. Conversely, in \cite{Farach1993}, the problem is approached in the following manner. For a given set of candidate measurement locations and a predefined number of measurements to be placed, find a measurement configuration that minimizes the error $\widehat{\bm{i}}_{u}-\bm{i}_{u}$. This approach follows the idea of using the covariance matrix to place measurements \cite{Schweppe1970, Schweppe1974}. The evaluation of all possible measurement configurations through a complete enumeration method is a formidable task. For example, to place 5 measurements in a system of 100 buses would require the evaluation of $\left(\frac{100}{5}\right)=75,287,520$ possible combinations; hence, a sequential solution process is adopted in \cite{Farach1993} under the assumption that the best $k+1$ measurement locations contain the best $k$ locations for all $k$. It turns out that this sequential scheme is not guaranteed to yield an optimal solution, as numerically demonstrated in \cite{Farach1993}. To circumvent this issue, a genetic algorithm is proposed in \cite{Kumar2005}. It is numerically demonstrated on small-scale test systems that the solution achieved using a genetic algorithm is optimal and hence superior to the sequential solution. 

The work in \cite{Madtharad2005} extends the methodology developed in \cite{Madtharad2003} to the specific application of power quality monitoring. Refer to Subsection \ref{sec.StaticStateEstimationT}, \emph{Algorithm 1}. See also \cite{Dag2012, Ketabi2012}.

\vspace{-.3cm}
\section{Measurement placement in distribution grids}\label{section.D}

\vspace{-.3cm}
In the U.S. electric power grid infrastructure, \emph{``there are approximately four times more low-voltage distribution substations than there are high-voltage substations''} \cite{NRC2009}; and each low-voltage distribution substation minimally houses one feeder that delivers power to a neighborhood---this implies that if represented by graphs, distribution grids certainly have at least 10 times as many nodes as transmission grids. This is a key difference between transmission and distribution grids, and it has an important effect on measurement placement. Note, however, that this difference in the number of nodes does not necessarily translate linearly into the technical requirements for (and cost of) measurement placement in distribution grids. It does, however, support the argument that a careful measurement placement is of paramount importance. Table \ref{tab.3} summarizes additional characteristics that distinguish distribution from transmission grids.

\begin{table}[ht!]
\centering \scriptsize
\setlength{\tabcolsep}{0.5em}
\caption{Distribution vs. transmission grids: Comparison of characteristics}
\begin{tabular}{l l l}
\cmidrule[1pt]{1-3}
\rowcolor{black!10}\textbf{Characteristic} & \textbf{Distribution grids} & \textbf{Transmission grids} \\ \hline
Topology & Radial to slightly meshed & Highly meshed \\ \hline
Phase unbalance/ & Significant/each phase & Negligible to moderate/ \\
circuit analysis & is analyzed individually & analysis of one phase \\ 
& & often suffices \\ \hline
Measurement system/ & RTU/unobservable, & RTU/observable, \\
observability & Micro-PMU/unobservable & PMU/unobservable\footnotemark[2] \\ \hline
Renewable generation/ & Dispersed/high/very high & Lumped at high-voltage \\
operational\footnotemark[8] uncertainty/ & & substations/moderate/ \\
planning\footnotemark[9] uncertainty & & high to very high \\ \hline
Load demand/ & Lumped at distribution & Lumped at high-voltage \\
operational uncertainty/ & transformers/high/ & substations/moderate/ \\
planning uncertainty & very high & high to very high \\ \hline
\cmidrule[1pt]{1-3}
\end{tabular}
\label{tab.3}
\end{table}
\footnotetext[2]{A few transmission grids in the United States are fully observable by PMUs \cite{Zhao2020}.}
\footnotetext[8]{Real time to a few hours ahead.}
\footnotetext[9]{Day ahead to weekly to yearly.}

Another aspect to consider is the system infrastructure: existing instrument transformers, availability of communications systems, space for expansion, etc. The nodes that represent a transmission grid always reside within a high-voltage substation, in which adequate infrastructure is often available. On the other hand, many nodes that represent a distribution grid are located in a section of an overhead (or underground) cable that crosses cities with no or minimal infrastructure. This imposes additional constraints on the candidate locations for measurement placement in distribution grids. These and other aspects are discussed next. As in the previous section on transmission, in this section, we elaborate on the optimal measurement placement problem as well as on challenges and solutions for distribution systems for several different applications, starting with static state estimation.

{\color{black}
\vspace{-.2cm}
\subsection{Static state estimation in distribution grids}\label{sec.StaticStateEstimationD}
}

Table \ref{tab.3} shows that, irrespective of the measurement system, distribution grids are not observable. The following remarks are in order.

\vspace{.2cm}
\begin{remark}[Observability in electric power transmission grids]
To this point, two definitions of observability are given: \underline{numerical observability} and \underline{topological observability}. See the box \emph{Observability for static state estimation in transmission grids} in Subsection \ref{sec.StaticStateEstimationT}. We refer to these definitions as strong notions of observability; they are widely used in transmission grids. 
\end{remark}

\vspace{-.6cm}
\begin{remark}[Observability in electric power distribution grids]
Static state estimation of distribution grids relies heavily on the use of pseudo-measurements \cite{Ghosh1997, Dehghanpour2019}, which are obtained from historical load data. Without the use of pseudo-measurements, distribution grids are not observable according to the strong notions of observability. 
\end{remark}

\vspace{-.6cm}
\begin{definition}
\underline{Weak numerical observability} is defined as the ability of the linear model (\ref{eq.sysmeasmodel}) to be solved for a state estimate $\widehat{\bm{x}}$, provided that the measurement vector $\bm{z}$ is augmented with pseudo-measurements.
\end{definition}

\vspace{-.2cm}
It is clear from the previous discussion that observability is yet a challenging concept to be applied in distribution grids. See \cite{Brinkmann2017}. For this reason, the placement of various measurement technologies---particularly when they are simultaneously considered---remains of high interest, and an optimal mix is contemplated in what follows. Note that this increases the complexity of the problem; whereas in Section \ref{section.T} the goal was to find the number and location of measurements, in this section the goal is to find the number, location, and \emph{type} of measurements. Table \ref{tab.4} summarizes the types of measurements considered in this subsection.

Also, in transmission grids, the algebraic state variables are defined as the voltage phasors (magnitude and phase angle) at each bus.\footnotemark[3]\footnotetext[3]{In certain cases, in addition to voltage phasors, the transformer taps and the firing angle of converters are also defined as algebraic state variables.} Accordingly, the static state estimation algorithms are designed to estimate the voltage phasors based on a set of measurements. Note that the choice of algebraic state variables is not unique. In distribution grids, two definitions of algebraic state variables are commonly used:

\vspace{-.1cm}
\begin{itemize}
\item Voltage phasors at each bus
\item Current phasors at each branch.
\end{itemize}

\vspace{-.1cm}
It is beyond the scope of this paper to discuss the advantages and disadvantages of these two options; see, e.g., \cite{Singh2009b, Primadianto2017}. Of importance here, however, is the following: Given that the measurement placement has a direct impact on the performance of the static state estimator \cite{Li1996, Wang2004}, the choice of algebraic state variables will affect the formulation of the measurement placement problem.

\begin{table*}[ht!]
\centering \scriptsize
\setlength{\tabcolsep}{0.1em}
\caption{Measurement placement methods for static state estimation of distribution grids (ordered by year of publication)}
\begin{tabular}{l ccccccccccccccccccccccccc}
\cmidrule[1pt]{1-26}
\rowcolor{black!10} 
\textbf{Meter (Measurand)}
& \cite{Baran1995}
& \cite{Baran1996} 
& \cite{Leou1996} 
& \cite{Shafiu2005}
& \cite{Muscas2006}
& \cite{Muscas2009}
& \cite{Singh2009}
& \cite{Singh2011}
& \cite{Nusrat2012}
& \cite{Liu2012}
& \cite{Pegoraro2013}
& \cite{Liu2014}
& \cite{Xiang2014}
& \cite{Damavandi2015}
& \cite{Chen2016}
& \cite{Xygkis2016}
& \cite{Brinkmann2017}
& \cite{Prasad2017}
& \cite{Xygkis2017}
& \cite{Prasad2018}
& \cite{Wang2018}
& \cite{Xygkis2018}
& \cite{Picallo2019}
& \cite{Teimourzadeh2019}
& \cite{Samudrala2020}
\\
Pseudo (kwatt, kvar)
& \checkmark 
& \checkmark 
& \checkmark 
& \checkmark 
& \checkmark 
& \checkmark 
& \checkmark 
& \checkmark 
& \checkmark 
& \checkmark 
& \checkmark 
& \checkmark 
& \checkmark 
& \checkmark 
& \checkmark 
& \checkmark 
& \checkmark 
& \checkmark 
& \checkmark 
& \checkmark 
& \checkmark 
& \checkmark 
& \checkmark 
& \checkmark 
& 
\\
\rowcolor{black!10}
Current (amp)
& \checkmark 
& \checkmark 
& \checkmark 
& 
& \checkmark 
& \checkmark 
& 
& 
& 
& 
& 
& 
& \checkmark 
& 
& \checkmark 
& 
& 
& 
& 
& 
& 
& 
& 
& 
& 
\\
Voltage (volt)
& 
& 
& \checkmark 
& \checkmark 
& 
& 
& \checkmark 
& \checkmark 
& \checkmark 
& \checkmark 
& \checkmark 
& \checkmark 
& \checkmark 
& \checkmark 
& \checkmark 
& 
& \checkmark 
& \checkmark 
& \checkmark 
& 
& \checkmark 
& \checkmark 
& 
& 
& \checkmark 
\\
\rowcolor{black!10} 
Power (kwatt, kvar)
& \checkmark 
& \checkmark 
& \checkmark 
& 
& \checkmark 
& \checkmark 
& \checkmark 
& \checkmark 
& \checkmark 
& \checkmark 
& \checkmark 
& \checkmark 
& 
& \checkmark 
& \checkmark 
& 
& 
& \checkmark 
& \checkmark 
& 
& \checkmark 
& \checkmark 
& 
& 
& (kwatt only) 
\\
PMU
& 
& 
& 
& 
& 
& 
& 
& 
& 
& \checkmark 
& 
& \checkmark 
& 
& \checkmark 
& 
& 
& 
& 
& 
& \checkmark 
& 
& 
& \checkmark 
& 
& 
\\
\rowcolor{black!10}
Micro-PMU
& 
& 
& 
& 
& 
& 
& 
& 
& 
& 
& 
& 
& 
& 
& 
& 
& 
& 
& 
& 
& 
& 
& 
& \checkmark 
& 
\\
Smart meter
& 
& 
& 
& 
& 
& 
& 
& 
& 
& \checkmark 
& 
& \checkmark 
& 
& 
& 
& \checkmark 
& 
& 
& 
& 
& 
& 
& 
& \checkmark 
& 
\\
\hline
\cmidrule[.75pt]{1-26}
\rowcolor{black!10}
\textbf{Algebraic state variables}
&&&&&&&&&&&&&&&&&&&&&&&&&
\\
Voltage phasors
& \checkmark 
& \checkmark 
& 
& \checkmark 
& 
& 
& \checkmark 
& \checkmark 
& \checkmark 
& \checkmark 
& \checkmark 
& \checkmark 
& \checkmark 
& \checkmark 
& \checkmark 
& \checkmark 
& \checkmark 
& \checkmark 
& \checkmark 
& \checkmark 
& \checkmark 
& \checkmark 
& \checkmark 
& \checkmark 
& 
\\
\rowcolor{black!10}
Current phasors
& 
& 
& \checkmark 
& 
& \checkmark 
& \checkmark 
& 
& 
& 
& 
& 
& 
& 
& 
& 
& 
& 
& 
& 
& 
& 
& 
& 
& 
& 
\\
\hline
\cmidrule[1pt]{1-26}
\end{tabular}
\label{tab.4}
\end{table*}

Having motivated the measurement placement problem and discussed its particularities to distribution grids, let us proceed to existing methods. 

\vspace{.2cm}
\noindent
\emph{Minimum number of measured quantities for weak observability:}

\vspace{.1cm}
A heuristic approach tailored to radial feeders is proposed in \cite{Baran1995, Baran1996}. A set of rules developed based on empirical observations is proposed to determine the number, location, and type of meters, as follows:

\vspace{-.3cm}
\begin{itemize}
\item Place a power meter at the substation.
\item Place current meters on all main switch and fuse locations that need to be monitored.
\item Place current meters along the feeder such that the total load in the zones defined by the meters are similar in magnitude.
\item Place current meters on all normally open tie switches used for feeder switching. Measurements of voltage magnitude at both ends of these tie switches are desirable.
\end{itemize}

\vspace{-.3cm}
The number of meters placed on the system if the previous rules are used alone might be prohibitively large. The authors circumvent this problem by adapting the method as follows \cite{Koglin1975}. First, the notion of \emph{interesting or influential quantities} is defined as any variable that can be expressed in terms of the algebraic state variables and is needed by monitoring and control applications, referred to as distribution automation functions. Then, the set of meters obtained with the previous rules is ranked based on their impact on the variance of the estimated interesting quantities. The ranking will indicate the order in which the meters are to be eliminated, if needed. This method represents the first reported attempt to place measurements in a principled fashion to aid in the static state estimation of distribution grids. Along similar lines, the work in \cite{Leou1996} relies on empirical observations to develop a rule-based heuristic approach aimed at reducing the variance of the estimated voltage magnitudes on those nodes that are not measured. As opposed to the deterministic approach taken in \cite{Baran1995, Baran1996}, uncertainties in the feeder loads, network parameters, and the calculated feeder node voltages are considered in \cite{Leou1996}; more specifically, the uncertainties are characterized by ranges of values---i.e., by intervals with confidence levels represented as fuzzy numbers---and the measurement placement scheme is assessed on all plausible system operating conditions. The rule-based approaches in \cite{Baran1995, Baran1996, Leou1996} (and their extensions) are simple but need to be adapted on a case-by-case basis, thereby requiring specific knowledge of the system. See also \cite{Liu2002}.

An important characteristic of modern distribution grids is the presence of distributed generators (DGs), which are not considered in the previous work. An endeavor to fulfill this gap is reported in \cite{Shafiu2005}, where the case study contemplates wind power plants connected to the 11-kV section of the U.K. generic distribution system\footnotemark[5]\footnotetext[5]{\url{https://github.com/sedg/ukgds}} (UKGDS); and in \cite{Muscas2006, Muscas2009}, where the case study contemplates portions of an Italian distribution grid to which wind, gas, and cogeneration power plants are connected. Following \cite{Leou1996}, these works rely on the variance of voltage \cite{Shafiu2005} or current \cite{Muscas2006, Muscas2009} magnitudes estimated for those nodes that are not measured as a metric to assess the quality of the obtained state estimation results. See also \cite{Nusrat2012, Brinkmann2017, Wang2018}.

The method in \cite{Shafiu2005} is heuristic and can be summarized as follows:
\vspace{-.3cm}
\begin{enumerate}
\item Define an initial set of buses, $\mathbb{S}$, where voltage meters are installed.
\item Solve a power flow algorithm for the peak load.
\item Solve a power flow algorithm for a random, $\pm 20\%$ of nominal load to consider uncertainties in load. Note that one can modify this algorithm to consider variations in net load---that is, load \emph{minus} variable distributed generation.
\item Calculate the difference: $e=\sum_{k=1}^{\text{card}(\mathbb{S})}\left(V_{k}^{peak}-V_{k}^{rand}\right)^2$.
\item If $e<$ preset threshold, store the power flow case obtained in 2.
\item Repeat 2--4 until $N$ (preset number of) power flow cases are obtained.
\item For each bus $b\notin\mathbb{S}$, calculate the $\widehat{\sigma}_{b}^{2}=\sqrt{\frac{\sum_{\ell=1}^{N}\left(V_{\ell}^{rand}-\text{mean}\left(V^{rand}\right)\right)^{2}}{N-1}}$.
\item If $\widehat{\sigma}_{b}^{2}<$ preset threshold for all $b\notin\mathbb{S}$, stop; else, continue.
\item Move the meters in the set $\mathbb{S}$ to buses with the largest variances $\widehat{\sigma}^{2}$.
\item If required, add a meter on the bus with the largest $\widehat{\sigma}^{2}$; update 1; go to 2. Note that the algorithm could be modified to remove meters from the initial list of buses if the variance is smaller under all loading conditions.
\end{enumerate}

\vspace{-.3cm}
The previous method is independent of any state estimator, as opposed to the method in \cite{Muscas2006, Muscas2009}, in which the problem definition is stated as follows. Find a measurement placement that makes the system \emph{observable with established accuracy} at a minimum cost. The accuracy requirement is a consequence of the use of pseudo-measurements with corresponding large uncertainties. For example, it is suggested in \cite{Singh2009} that the error in the pseudo-measurements is in between 20\% and 50\%. Formally, consider a system with $n$ state variables, and define $\sigma_{max}^{2}$, the maximum acceptable variance of any estimated state variable. Then, for a given set of measurements:

\vspace{-.3cm}
\begin{equation}\label{eq.41x}
\begin{aligned}
\min_{} \quad & J = \sum_{k=1}^{n} \left(\frac{\sigma_{k}}{\sigma_{max}}\right)^{2}, \\
\textrm{s.t.} \quad & \sigma_{k} \le \sigma_{max} \quad \forall\, k=1,...,n,
\end{aligned}
\end{equation}

\noindent
where $\sigma_{k}$ is the variance associated with the estimated state variable, $\widehat{x}_{k}$. Finding the optimal solution to the combinatorial optimization problem in (\ref{eq.41x}) is challenging. A suboptimal solution to (\ref{eq.41x}) is achieved in \cite{Muscas2006, Muscas2009} by using dynamic programming. The proposed method proceeds as follows. First, using the model of a distribution grid of interest:

\vspace{-.2cm}
\begin{itemize}
\item Build sets of true values for $N$ operating conditions by solving a power flow algorithm for different random loading scenarios.
\item For each operating condition, build sets of \emph{synthetic measurements} by adding random noise to the power flow variables.
\end{itemize}

\vspace{-.2cm}
Next, assume that measurement devices are installed at the substation and DG buses. Accordingly, synthetic measurements corresponding to these locations are used in an attempt to solve (\ref{eq.41x}). If a solution is found, no additional measurements are required. Else, define a set $\mathbb{C}$ of candidate locations for measurement placement. Also, define:

\vspace{-.2cm}
\begin{equation}
\epsilon=\frac{1}{N}\sum_{\ell=1}^{N}J_{\ell}.
\end{equation}

\vspace{-.1cm}
Then, beginning from the initial configuration with measurements at the substation and DG locations:

\vspace{-.2cm}
\begin{enumerate}
\item Evaluate $\epsilon$ for all candidate locations in $\mathbb{C}$.
\item Place a measurement at the candidate location that yields min$(\epsilon)$.
\item Remove the corresponding candidate location from $\mathbb{C}$.
\item If min$(\epsilon)<$ preset value and the constraints in (\ref{eq.41x}) are not violated, stop; else, go to 1.
\end{enumerate}

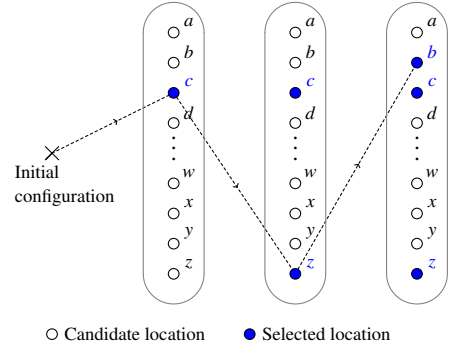
\begin{figure}
\centering
\definecolor{yqyqyq}{rgb}{0.5019607843137255,0.5019607843137255,0.5019607843137255}
\begin{tikzpicture}[line cap=round,line join=round,>=triangle 45,x=1cm,y=1cm,scale=1]
\clip(7.0,1.65) rectangle (12.9,6.25);
\draw [line width=0.4pt,dash pattern=on 1pt off 1pt] (7.6,4.2)-- (9.2,5);
\draw [line width=0.4pt] (8.452089023958912,4.626044511979456) -- (8.44340751996576,4.57829624001712);
\draw [line width=0.4pt] (8.452089023958912,4.626044511979456) -- (8.408681503993153,4.647748271962338);
\draw [line width=0.4pt,dash pattern=on 1pt off 1pt] (9.2,5)-- (10.8,2.6);
\draw [line width=0.4pt] (10.032304241308774,3.7515436380368397) -- (9.983847879345612,3.7542356581459044);
\draw [line width=0.4pt] (10.032304241308774,3.7515436380368397) -- (10.04845636196316,3.797307979890936);
\draw [line width=0.4pt,dash pattern=on 1pt off 1pt] (10.8,2.6)-- (12.4,5.4);
\draw [line width=0.4pt] (11.62889379181119,4.050564135669583) -- (11.648156319685317,4.006019539960664);
\draw [line width=0.4pt] (11.62889379181119,4.050564135669583) -- (11.580737472125874,4.044544595708919);
\draw (7.0,4.15390317111036) node[anchor=north west] {\parbox{1.0 cm}{\scriptsize Initial\\configuration}};
\draw [line width=0.4pt,color=yqyqyq] (8.8,5.8)-- (8.8,2.6);
\draw [line width=0.4pt,color=yqyqyq] (9.6,5.8)-- (9.6,2.6);
\draw [shift={(9.2,5.8)},line width=0.4pt,color=yqyqyq] plot[domain=0:3.141592653589793,variable=\t]({1*0.4*cos(\t r)+0*0.4*sin(\t r)},{0*0.4*cos(\t r)+1*0.4*sin(\t r)});
\draw [shift={(9.2,2.6)},line width=0.4pt,color=yqyqyq] plot[domain=3.141592653589793:6.283185307179586,variable=\t]({1*0.4*cos(\t r)+0*0.4*sin(\t r)},{0*0.4*cos(\t r)+1*0.4*sin(\t r)});
\draw [line width=0.4pt,color=yqyqyq] (10.4,5.8)-- (10.4,2.6);
\draw [line width=0.4pt,color=yqyqyq] (11.2,5.8)-- (11.2,2.6);
\draw [line width=0.4pt,color=yqyqyq] (12,5.8)-- (12,2.6);
\draw [line width=0.4pt,color=yqyqyq] (12.8,5.8)-- (12.8,2.6);
\draw [shift={(10.8,5.8)},line width=0.4pt,color=yqyqyq] plot[domain=0:3.141592653589793,variable=\t]({1*0.4*cos(\t r)+0*0.4*sin(\t r)},{0*0.4*cos(\t r)+1*0.4*sin(\t r)});
\draw [shift={(12.4,5.8)},line width=0.4pt,color=yqyqyq] plot[domain=0:3.141592653589793,variable=\t]({1*0.4*cos(\t r)+0*0.4*sin(\t r)},{0*0.4*cos(\t r)+1*0.4*sin(\t r)});
\draw [shift={(12.4,2.6)},line width=0.4pt,color=yqyqyq] plot[domain=3.141592653589793:6.283185307179586,variable=\t]({1*0.4*cos(\t r)+0*0.4*sin(\t r)},{0*0.4*cos(\t r)+1*0.4*sin(\t r)});
\draw [shift={(10.8,2.6)},line width=0.4pt,color=yqyqyq] plot[domain=3.141592653589793:6.283185307179586,variable=\t]({1*0.4*cos(\t r)+0*0.4*sin(\t r)},{0*0.4*cos(\t r)+1*0.4*sin(\t r)});
\draw [color=black](9.05,4.75) node[anchor=north west] {$\vdots$};
\draw [color=black](10.65,4.75) node[anchor=north west] {$\vdots$};
\draw [color=black](12.25,4.75) node[anchor=north west] {$\vdots$};
\draw (7.65,2) node[anchor=north west] {\scriptsize Candidate location};
\draw (10.25,2) node[anchor=north west] {\scriptsize Selected location};
\begin{scriptsize}
\draw [fill=blue] (10.2,1.8) circle (2pt);
\draw [color=black] (7.6,1.8) circle (2pt);
\draw [color=black] (9.2,5.8) circle (2pt);
\draw[color=black] (9.40,5.937044425948331) node {$a$};
\draw [color=black] (9.2,5.4) circle (2pt);
\draw[color=black] (9.40,5.540790813762116) node {$b$};
\draw [fill=blue] (9.2,5) circle (2pt);
\draw[color=blue] (9.40,5.139982562355368) node {$c$};
\draw [color=black] (9.2,4.6) circle (2pt);
\draw[color=black] (9.40,4.739174310948622) node {$d$};
\draw [color=black] (9.2,3.8) circle (2pt);
\draw[color=black] (9.40,3.9375578081351277) node {$w$};
\draw [color=black] (9.2,3.4) circle (2pt);
\draw[color=black] (9.40,3.5367495567283807) node {$x$};
\draw [color=black] (9.2,3) circle (2pt);
\draw[color=black] (9.40,3.140495944542165) node {$y$};
\draw [color=black] (9.2,2.6) circle (2pt);
\draw[color=black] (9.40,2.739687693135418) node {$z$};
\draw [color=black] (7.6,4.2)-- ++(-2.5pt,-2.5pt) -- ++(5pt,5pt) ++(-5pt,0) -- ++(5pt,-5pt);
\draw [color=black] (10.8,5.8) circle (2pt);
\draw[color=black] (11.00,5.937044425948331) node {$a$};
\draw [color=black] (12.4,5.8) circle (2pt);
\draw[color=black] (12.60,5.937044425948331) node {$a$};
\draw [color=black] (10.8,5.4) circle (2pt);
\draw[color=black] (11.00,5.540790813762116) node {$b$};
\draw [fill=blue] (12.4,5.4) circle (2pt);
\draw[color=blue] (12.60,5.540790813762116) node {$b$};
\draw [color=black] (10.8,4.6) circle (2pt);
\draw[color=black] (11.00,4.739174310948622) node {$d$};
\draw [color=black] (12.4,4.6) circle (2pt);
\draw[color=black] (12.60,4.739174310948622) node {$d$};
\draw [fill=blue] (10.8,5) circle (2pt);
\draw[color=blue] (11.00,5.139982562355368) node {$c$};
\draw [fill=blue] (12.4,5) circle (2pt);
\draw[color=blue] (12.60,5.139982562355368) node {$c$};
\draw [color=black] (10.8,3.8) circle (2pt);
\draw[color=black] (11.00,3.9375578081351277) node {$w$};
\draw [color=black] (12.4,3.8) circle (2pt);
\draw[color=black] (12.60,3.9375578081351277) node {$w$};
\draw [color=black] (10.8,3.4) circle (2pt);
\draw[color=black] (11.00,3.5367495567283807) node {$x$};
\draw [color=black] (12.4,3.4) circle (2pt);
\draw[color=black] (12.60,3.5367495567283807) node {$x$};
\draw [color=black] (10.8,3) circle (2pt);
\draw[color=black] (11.00,3.140495944542165) node {$y$};
\draw [color=black] (12.4,3) circle (2pt);
\draw[color=black] (12.60,3.140495944542165) node {$y$};
\draw [fill=blue] (10.8,2.6) circle (2pt);
\draw[color=blue] (11.00,2.739687693135418) node {$z$};
\draw [fill=blue] (12.4,2.6) circle (2pt);
\draw[color=blue] (12.60,2.739687693135418) node {$z$};
\end{scriptsize}
\end{tikzpicture}
\caption{Pictorial idea of the sequential method in \cite{Muscas2006, Muscas2009}.}
\label{fig.3x}
\end{figure}

\vspace{-.2cm}
The procedure is illustrated in Fig. \ref{fig.3x}, where $\mathbb{C}=\{a,b,...,z\}$. Starting from the initial configuration, $\epsilon$ is evaluated for all candidate locations, and $c$ yields min$(\epsilon)$. An additional measurement is placed at this location, but the stopping criteria are not met; thus, $\epsilon$ is reevaluated for each of the remaining candidate locations, and $z$ yields min$(\epsilon)$. This process is repeated again before the stopping criteria are met. At the end, three additional measurements are placed at locations $\{b,c,z\}$. It is reported in \cite{Muscas2009} that approximately 29,000 combinations are evaluated by this method before a suboptimal solution to (\ref{eq.41x}) is achieved for a distribution grid of 51 nodes. For the same grid, an optimal solution to (\ref{eq.41x}) using a complete enumeration method would require the evaluation of combinations in the order of $10^{13}$. The algorithm developed in \cite{Muscas2006, Muscas2009} is further exploited in \cite{Pegoraro2013} with an extended error covariance matrix that accounts for network model parameter uncertainties.

An extension of the previous methods is developed in \cite{Singh2009, Singh2011} in which the variance of voltage magnitudes \emph{and phase angles}, estimated on those nodes that are not measured, is used as a metric to evaluate the obtained results. The authors start by defining:

\vspace{-.4cm}
\begin{equation}\label{eq.42x}
p_{k} = \text{Pr}\left\{\left|\frac{|\widehat{v}_{k}|-|v_{k}|}{|v_{k}|}\right|<\epsilon_{v}, \left|\frac{\widehat{\theta}_{k}-\theta_{k}}{\theta_{k}}\right|<\epsilon_{_{\theta}} \right\}, \quad \forall\, k=2,...,n_{bu},
\end{equation}

\vspace{-.2cm}
\noindent
where $\epsilon_{v}$, $\epsilon_{_{\theta}}$ are predefined thresholds. The goal is to obtain a probability index $p_{k}>0.95$ for all but the substation bus, $k=1$, where supposedly an accurate measurement device is already in place. From (\ref{eq.42x}):

\vspace{-.4cm}
\begin{align}
\bm{\mu}_{k} &= \mathcal{E}
\left(
\left[
\begin{array}{c}
|\widehat{v}_{k}| \\
\widehat{\theta}_{k}
\end{array}
\right]
\right), \\
\bm{R}_{k} &= \mathcal{E}
\left(
\left(
\left[
\begin{array}{c}
|\widehat{v}_{k}| \\
\widehat{\theta}_{k}
\end{array}
\right]
-\bm{\mu}_{k}
\right)
\left(
\left[
\begin{array}{c}
|\widehat{v}_{k}| \\
\widehat{\theta}_{k}
\end{array}
\right]
-\bm{\mu}_{k}
\right)\tran 
\right).\label{eq.44x}
\end{align}

\vspace{-.2cm}
Note that the error covariance matrix, $\bm{R}_{k}$, associated with bus $k$ is two-dimensional. The authors rely on the geometric interpretation of the error covariance matrix---in particular, the fact that the two-dimensional error covariance matrix can be geometrically seen as an ellipse---as an indication to where measurements shall be allocated. The strategy is to place measurements at the locations where the ellipse area, proportional to $\sqrt{\det\bm{R}_{k}}$, is largest. The accuracy of the estimated state variables is improved by shrinking the ellipse areas. Ultimately, this is a way to reduce the variance of the estimates, as in previous work. But this idea is nicely cast on a more formal statistical foundation in \cite{Singh2009, Singh2011}. The algorithm in \cite{Singh2009} is as follows:

\vspace{-.2cm}
\begin{enumerate}
\item Run the state estimator over a set of Monte Carlo simulations.
\item For $\epsilon_{v}=1\%$, $\epsilon_{_{\theta}}=5\%$, if $p_{k}>95\%$, $k=2,...,n_{bu}$, stop; else go to 3.
\item If the relative errors in voltage magnitude are within bounds, go to 5; else go to 4.
\item Place a voltage meter on bus $k$ with the largest $\sqrt{\det\bm{R}_{k}}$.
\item Compute the error covariance matrix corresponding to the real and reactive power flow in each branch using the accordingly modified (\ref{eq.44x}).
\item Place a power meter on branch $\ell$ with the largest $\sqrt{\det\bm{R}_{\ell}}$; go to 1.
\end{enumerate}

\vspace{-.2cm}
The algorithm proposed in \cite{Singh2011} extends and improves on the previous algorithm. See also \cite{Prasad2017, Prasad2018}.

Following previous work, the variance of the estimated quantities is the adopted criterion in \cite{Chen2016} to evaluate the estimation accuracy; however, the novelty in \cite{Chen2016} is to link the gain matrix to a circuit representation. For example, nodal voltage measurements are represented by shunt admittances---branch current, power flow, and power injection measurements are also considered. The gain matrix is represented analytically in the form of a network admittance matrix, which enables the reformulation of the measurement placement problem as a mixed-integer linear programming problem with disjunctive inequalities.

\vspace{.2cm}
\noindent
\emph{Minimum number of measurement points for weak observability:}

\vspace{.1cm}
In all the previous work, the goal is to minimize the total number of measured quantities, i.e., the total number of measured voltages, currents, and powers. The authors of \cite{Xiang2014} argue that minimizing the total number of measured quantities across the grid might lead to a large number of geographically spread measured points, thereby leading to the need for sensors at more locations and the resulting overall high cost. Alternatively, they advocate that the optimization could focus on ensuring the maximum utilization of available information from a few locations, i.e., it might be cost-effective to measure as many quantities as possible from the same or a few locations; therefore, the goal here is to minimize the total number of measured points while maximizing the knowledge from each of those points---a goal that was also pursued by \cite{Muscas2009} in identifying the use of more current measurements from the same node. Apart from this philosophical difference, the measurement placement method proposed in \cite{Xiang2014} follows the same lines of previous works, as summarized next:

\vspace{.1cm}
\emph{For all medium-voltage substations:}

\vspace{-.3cm}
\begin{itemize}
\item Place voltage measurements on each bus.
\item Place current measurements on each feeder leaving the substation.
\end{itemize}

\vspace{-.2cm}
\emph{For all large industrial loads and DGs:}

\vspace{-.3cm}
\begin{itemize}
\item Place voltage measurements on each large industrial load and DG bus.
\item Place current measurements on each branch connected to large industrial loads and DG buses.
\end{itemize}

\vspace{-.2cm}
Further, at least one point of connection with household loads and one point of connection with commercial loads must be measured following the scheme for large industrial loads and DGs. Finally, following previous work, additional measurements should be placed to improve the performance of the static state estimator; however, previous works use the variance of voltage magnitudes and phase angles, estimated on those nodes that are not measured, as a metric to evaluate the obtained results. Conversely, \cite{Xiang2014} uses the variance of branch voltage phasors as a primary criterion and the variance of complex power flows as a secondary criterion. The overall strategy in \cite{Xiang2014} seems to be supported by a Dutch distribution grid operator and is well justified; the trade-offs between the proposed method and previous work are not elaborated.

\vspace{.2cm}
\noindent
\emph{Consideration of advanced metering infrastructure:}

\vspace{.1cm}
To this point, standard meters---that is, current, voltage, and power meters---have been considered. But the next generation of distribution grids will benefit from more advanced measurement systems \cite{Heydt2010}, referred to as advanced metering infrastructure. Accordingly, a first attempt to formulate the measurement placement problem for distribution grids to consider PMUs and smart meters, in addition to standard meters, is reported in \cite{Liu2012}. The problem is formulated as an optimization, and it follows closely the idea in \cite{Singh2009}. Formally:

\vspace{-.6cm}
\begin{equation}\label{eq.45x}
\begin{aligned}
\min_{} \quad & \sum_{k=1}^{n_{bu}}c_{k}^{pmu} \cdot u_{k}^{pmu} + \sum_{k=1}^{n_{bu}}c_{k}^{sm} \cdot u_{k}^{sm} + \beta_{v}\sum_{\ell=1}^{N}e_{v}^{\ell} + \beta_{_{\theta}}\sum_{\ell=1}^{N}e_{_{\theta}}^{\ell}, \\
\textrm{s.t.} \quad &
\begin{cases}
e_{v}^{\ell} \le \epsilon_{v}, \\
e_{_{\theta}}^{\ell} \le \epsilon_{_{\theta}}, \quad \forall\,\ell=1,...,N,
\end{cases}
\end{aligned}
\end{equation}

\vspace{-.2cm}
\noindent
where $c_{k}^{pmu}$ $\left(c_{k}^{sm}\right)$ denotes the cost of installing a PMU (smart meter) on bus $k$; $\beta_{v}$ and $\beta_{_{\theta}}$ are parameters that can be used to give different weights to voltage amplitude and phase deviations; and $N$ is the number of considered operating conditions;

\vspace{-.4cm}
\begin{equation}
u_{k}^{pmu} \left[u_{k}^{sm}\right] =
\begin{cases}
1 \quad \text{if a PMU [smart meter] is installed on bus } k,\\
0 \quad \text{otherwise},
\end{cases}
\end{equation}

\vspace{-.2cm}
\begin{equation}
e_{v}^{k} := \max_{k}\left|\frac{|\widehat{v}_{k}|-|v_{k}|}{|v_{k}|}\right|; \quad \text{and} \quad e_{_{\theta}}^{k} := \max_{k}\left|\widehat{\theta}_{k}-\theta_{k}\right|.
\end{equation}

The optimization problem (\ref{eq.45x}) is solved by a genetic algorithm. Solutions that violate accuracy limits are penalized, thereby increasing their cost. Solutions with the same costs are ranked by their accuracy. Finally, the method finds a Pareto optimal front that finds a measurement placement solution for varying degrees of placement costs and estimation accuracy under several network configurations. This approach in \cite{Liu2012} is extended in \cite{Liu2014} to consider the uncertainties associated with DGs. Specifically, the DG outputs are used as pseudo-measurements with (non-Gaussian) unknown probability distribution functions modeled as Gaussian mixture models.

The work in \cite{Damavandi2015} develops a PMU placement method that extends the concept in \cite{Li2011} to distribution grids. The trace of $\bm{G}^{-1}$ (refer to A-optimality in Subsection \ref{subsec.AccuracyT}) is the selected criterion to evaluate the estimation accuracy. Following previous work \cite{Muscas2006, Muscas2009, Singh2009, Singh2011, Liu2012, Nusrat2012, Pegoraro2013, Liu2014, Xiang2014, Chen2016}, the authors also resort to Monte Carlo simulations; this is because measurement placement methods for distribution systems must account for frequent topological reconfigurations. The formulated problem is solved by using a robust submodular optimization algorithm, referred to as a submodular saturation algorithm. It is demonstrated through numerical simulations that the submodular saturation algorithm outperforms greedy and genetic algorithms in most cases. See also \cite{Teimourzadeh2019}, which focuses specifically on microgrids. Following \cite{Damavandi2015}, the works in \cite{Xygkis2016, Xygkis2017, Xygkis2018} also approach the measurement placement problem from the standpoint of an optimal experimental design; only standard meters are considered. The largest diagonal entry of $\bm{G}^{-1}$ is the selected (M-optimality) criterion to evaluate the estimation accuracy in \cite{Xygkis2016, Xygkis2017}, whereas a D-optimality criterion is selected in \cite{Xygkis2018}. Finally, the work in \cite{Picallo2019} explores properties such as the convexity and the modularity of different metrics in the context of an optimal experimental design to propose and compare several tight lower and upper bounds on the performance of the optimal solution; the focus is exclusively on the placement of PMUs.

Measurement placement heuristics/formulations for static state estimation in distribution grids rely heavily on pseudo-measurements, whether or not the notion of weak numerical observability is considered. Though convenient to address short-term goals, this strong dependence on pseudo-measurements is not effective as a long-term measurement placement strategy. This is further aggravated by distributed generation that adds another layer of uncertainty on pseudo-measurement models. The progressive elimination of high-uncertainty pseudo-measurements via a multistage formulation is a gap in the literature and an opportunity for future development. This is further discussed in Section \ref{section.discussion}.

{\color{black}
\vspace{-.2cm}
\subsection{Stability assessment and control in distribution grids}\label{sec.StabilityAssessmentAndControlD}
}

Given that distribution system state estimation is still an emerging field, some studies have considered placing measurements of power to improve load flow calculations and the consequent voltage control strategies. A method for placing power flow measurements in low-voltage networks to improve power flow calculations in medium-voltage networks is proposed in \cite{Cataliotti2016a, Cataliotti2016b}. Along similar lines, but with a focus on volt-var control, \cite{Zamani2018} proposes a measurement placement strategy for identifying the most important locations that can help to achieve the best performance of conservation voltage reduction. The algorithm starts with an initial set of meters and estimates the voltage profiles. Using the standard deviation of estimated nodal voltages, the least significant measurement is removed based on the voltage estimation accuracy, until one is left with the most significant locations for voltage measurements.

Stability assessment and associated control schemes represent a new paradigm in distribution grids. Accordingly, the number of works on measurement placement with this focus is scarce. This scenario is rapidly changing with the integration of DERs to modern distribution grids. Except for microgrids operating in island mode, the focus on stability assessment and control for measurement placement in distribution grids remains less relevant.

\vspace{.2cm}
\subsection{Topology change detection in distribution grids}\label{sec.TopologyChangeDetectionD}
Most previous work addresses the high degree of uncertainty in power distribution grids by relying on Monte Carlo simulations. An interesting departure from this trend is found in \cite{Samudrala2020}. The authors recognize that the probabilistic approach based on Monte Carlo simulations depends on the statistics of measurement noise and pseudo-measurements, which are unknown and time-varying. Instead, the authors propose a deterministic approach based on grid structural notions---namely, \emph{topology detectability} and \emph{outage identifiability}---that depend only on the system topology under normal operating conditions. Note, however, that distributed generation is not considered. The goal in \cite{Samudrala2020} is to find a measurement configuration to guarantee that topology changes can be detected and identified. The problem is formulated as an optimization problem that ensures topology change detection at minimal cost. Given the definitions:

\vspace{-.3cm}
\begin{itemize}
\item $\mathbb{B}$: the set of all buses in the distribution grid
\item $\mathbb{B}_{k}$: the set of all buses downstream of bus $k$ in a radial network
\item $\mathbb{E}$: the set of all branches in the distribution grid
\item $\mathbb{I}_{0}$: the set of all zero-injection buses in the distribution grid
\item $c_{k}^{bu}$: the cost of installing a meter on bus $k$
\item $c_{(k,\ell)}^{br}$: the cost of installing a meter on branch $(k,\ell)$
\item $u_{k}^{bu}=1$ if a meter is installed on bus $k$, and $0$ otherwise
\item $u_{k}^{br}=1$ if a meter is installed on branch $(k,\ell)$, and $0$ otherwise
\item $d_{k}$: the degree of bus $k$, i.e., number of branches incident to bus $k$
\item $r_{k}$: the index of the bus immediately upstream of bus $k$, i.e., the parent bus.
\end{itemize}

\vspace{-.3cm}
Then:

\vspace{-.4cm}
\begin{equation}\label{eq.46xxx}
\begin{aligned}
\min_{\bm{u}^{bu},\,\bm{u}^{br}} \quad & \sum_{k\in\mathbb{B}}c_{k}^{bu}u_{k}^{bu} + \sum_{(k,\ell)\in\mathbb{E}}c_{(k,\ell)}^{br}u_{(k,\ell)}^{br}, \\
\textrm{s.t.} 
\quad & d_{1}u_{1}^{bu} + \sum_{\ell\in \mathbb{B}_{1}}u_{\ell}^{bu} + \sum_{(1,\ell)\in\mathbb{E}}u_{(1,\ell)}^{br}\ge d_{1}-1, \\
\quad & d_{q}u_{q}^{bu} + \sum_{\ell\in \mathbb{B}_{q}}u_{\ell}^{bu} + \sum_{(q,\ell)\in\mathbb{E}}u_{(q,\ell)}^{br} \\
\quad & \qquad\ge d_{q}-2 \quad\forall\; q \in\mathbb{B}\backslash\{1\} \;\text{having}\; d_{q}\ge 3, \\
\quad & u_{q}^{bu} + u_{(r_{q},q)}^{br}=1 \quad\forall\; q\in\mathbb{I}_{0}, \\
\quad & u_{\ell}^{bu} + u_{(k,\ell)}^{br}\le 1 \quad\forall\; (k,\ell)\in\mathbb{E}.
\end{aligned}
\end{equation}

\vspace{-.1cm}
It is proven in \cite{Samudrala2020} that the constraints in (\ref{eq.46xxx}) guarantee topology change identifiability; the solution to (\ref{eq.46xxx}) is obtained by using a dynamic programming algorithm. We remark that \cite{Picallo2019, Samudrala2020} are, to the best of our knowledge, the only published works that present numerical results on distribution test systems of relevant size. See Table \ref{tab.dist.works}.

\vspace{-.2cm}
\subsection{Fault detection in distribution lines}\label{sec.FaultDetectionD}

\vspace{-.2cm}
The work in \cite{Pereira2004} is the first to consider the problem of measurement placement---specifically PMUs---for fault location in distribution grids. Initially, one PMU is placed at the substation. The prefault and fault-on current phasors acquired by the PMU at the substation are used to calculate the fault resistance and the fault-on voltage phasors. Then, additional PMUs scattered at strategic locations provide fault-on voltage phasor measurements. These locations are chosen to minimize the error between the calculated and measured voltage phasors. Formally:

\vspace{-.2cm}
\begin{equation}\label{eq.46x}
\begin{aligned}
\min_{} \quad & \sum_{k=1}^{n_{bu}}\sum_{\substack{ \ell=1 \\ \ell\ne k}}^{n_{bu}} u_{k\ell} \\
\textrm{s.t.} 
\quad &
\begin{cases}
u_{k\ell}=1, \quad \text{if } \sum_{q=1}^{n_{c}} \left|v_{q}^{(k)} - v_{q}^{(\ell)} \right| = 0, \\
u_{k\ell}=0, \quad \text{otherwise,}
\end{cases}
\end{aligned}
\end{equation}

\vspace{-.2cm}
\noindent
where:

\vspace{-.2cm}
\begin{itemize}
\item $n_{c}$: number of candidate locations for measurement placement
\item $v_{q}^{(k)}$: measured voltage phasor on bus $q$ for a fault at bus $k$
\item $v_{q}^{(\ell)}$: calculated voltage phasor on bus $q$ for a fault at bus $\ell$.
\end{itemize}

\vspace{-.2cm}
The optimization problem (\ref{eq.46x}) is solved by using a Tabu search \cite{Pereira2004}. This method is extended in \cite{Biscaro2010} to account for multiple fault scenarios by using Monte Carlo simulations. In this case, instead of a Tabu search, a greedy randomized adaptive search metaheuristic is adopted. Notice that these methods \cite{Pereira2004, Biscaro2010} are specifically designed to algorithms that use voltage sag information to locate faults.

The work in \cite{Jamei2018} is mostly concerned with the detection of anomalies in distribution grids using micro-PMUs. A comprehensive anomaly detection framework is developed. The authors recognize that it is not feasible to deploy micro-PMUs on all system buses and that the performance of the developed framework depends on the number and location of available micro-PMUs; therefore, a micro-PMU placement methodology specifically tailored to the anomaly detection framework is developed as follows. For a three-phase system of $n_{bu}$ buses, the following algebraic equation holds true during steady-state operation:

\vspace{-.2cm}
\begin{equation}\label{eq.48x}
\bm{L}\bm{d} = \bm{0}.
\end{equation}

\vspace{-.1cm}
\noindent
where:

\vspace{-.2cm}
\begin{equation}
\bm{L} =
\left(\bm{I}_{3n_{bu}} | -\bm{Y}\right); \quad
\bm{d} =
\left[\bm{i}_{1}\tran \;\; \bm{i}_{2}\tran \;\; \bm{i}_{3}\tran \;\; \bm{v}_{1}\tran \;\; \bm{v}_{2}\tran \;\; \bm{v}_{3}\tran \right]\tran ;
\end{equation}

\noindent
$\bm{I}_{3n_{bu}}$ is the identity matrix; $\bm{Y}$ is the three-phase admittance matrix; and $\bm{i}_{1}\tran$ ($\bm{v}_{1}\tran$) is a vector of the bus current injection (bus voltage) phasors, associated with phase number $1$. The symbol ``|'' is used to denote that the matrix $\bm{L}$ is composed by two sub-matrices; the top sub-matrix is $\bm{I}_{3n_{bu}}$, and the bottom sub-matrix is $-\bm{Y}$. Now, define a transformation matrix $\bm{T}$ such that:

\vspace{-.2cm}
\begin{equation}
\bm{T}=
\left[\begin{array}{cc}
\bm{T}_{u} \\
\bm{T}_{a}
\end{array}\right] \rightarrow
\bm{T}\bm{d}=
\left[\begin{array}{cc}
\bm{d}_{u} \\
\bm{d}_{a}
\end{array}\right],\;
\bm{L}\bm{T}\tran =
\left(\bm{L}_{u} | \bm{L}_{a}\right),
\end{equation}

\vspace{-.1cm}
\noindent
where $\bm{d}_{a}$ and $\bm{d}_{u}$ contain variables associated with buses in which micro-PMUs are available and unavailable, respectively. It follows that:

\vspace{-.3cm}
\begin{equation}\label{eq.50x}
\bm{L}_{u}\bm{d}_{u} = - \bm{L}_{a}\bm{d}_{a}.
\end{equation}

\vspace{-.05cm}
The key idea in \cite{Jamei2018} is to project (\ref{eq.50x}) onto the subspace spanned by the left singular vector, $\bm{\upsilon}_{u}$, corresponding to the smallest singular value of $\bm{L}_{u}$. It is expected that this procedure suppresses the effect of quantities associated with buses in which micro-PMUs are not available, thereby making the anomaly detection framework a function of available micro-PMU measurements only. In other words, in steady-state operation, $\bm{\upsilon}_{u}\hermconj\bm{L}_{u}\bm{d}_{u}$ should be numerically small, and the normalized quantity:

\vspace{-.3cm}
\begin{equation}\label{eq.63z}
\frac{\left|\bm{\upsilon}_{u}\hermconj\bm{L}_{a}\bm{d}_{a}\right|}{\norm{\bm{d}_{a}}^{2}},
\end{equation}

\vspace{-.05cm}
\noindent
should vary smoothly in time. Consequently, anomalies are detected from abrupt variations in (\ref{eq.63z}). Accordingly, the micro-PMU placement is formulated as a min-max optimization problem, as follows:

\vspace{-.5cm}
\begin{equation}\label{eq.52x}
\begin{aligned}
\bm{\Pi}^{\text{opt}} = \min_{\bm{\Pi}} \quad & \max_{\bm{d}_{a}}\frac{\bm{d}_{a}\hermconj\bm{X}\bm{d}_{a}}{\norm{\bm{d}_{a}}^{2}}, \\
\textrm{s.t.} \quad & \left(\bm{L}_{u}|\bm{L}_{a}\right)=\bm{L}\left(\bm{T}_{u}\tran |\bm{T}_{a}\tran \right), \\
\quad & \bm{T} = \bm{I}_{2}\otimes\left(\bm{\Pi}\otimes\bm{I}_{3}\right), \\
\quad & \bm{X} = \bm{L}_{a}\hermconj\bm{\upsilon}_{u}\bm{\upsilon}_{u}\hermconj\bm{L}_{a}, \\
\quad & \left[\bm{\Pi}\right]_{k\ell} \in \{0,1\},\; \sum_{k}\left[\bm{\Pi}\right]_{k\ell}=1,\; \sum_{\ell}\left[\bm{\Pi}\right]_{k\ell}=1.
\end{aligned}
\end{equation}

\vspace{-.3cm}
The formulation in (\ref{eq.52x}) seeks to minimize the number of measurement devices while maximizing the range of buses where anomalies can be detected. The min-max optimization problem (\ref{eq.52x}) is solved by using a greedy search. This approach has been recently extended to the problem of PMU placement for fault location \cite{Jamei2020}.

\vspace{-.2cm}
\subsection{Power quality monitoring in distribution grids}\label{sec.PowerQualityMonitoringD}

\vspace{-.2cm}
A measurement placement method for power quality estimation based on the notion of entropy is proposed in \cite{Ali2016}. Monte Carlo simulations are used to draw samples of network states at metered and nonmetered locations. Then, Bayesian inference is used to obtain maximum-likelihood estimates of the states at nonmetered locations. Based on simulated and estimated states at nonmetered locations, the most poorly predicted nonmetered locations are selected as the location for the next meter. See also \cite{Ma1996}.

\vspace{-.3cm}
\section{Discussion and avenues for future research}\label{section.discussion}

\vspace{-.3cm}
Previous sections revisited and discussed the different methods for measurement placement in transmission and distribution grids. The methods were categorized by application. To this point, the paper provides a fairly comprehensive description of the critical factors that go into formulating a measurement placement problem. Although state estimation remains the central application, the paper also summarizes the use of measurement placement algorithms for other applications related to power system stability and online security assessment. Now, the objective of this section is to provide a summary of what was discussed in previous sections with a forward-looking view.

Allocation algorithms developed for state estimation problems generally defined observability in a numerical or topological sense, and past works typically tried to optimize the sensor allocation to either maximize the observability against all uncertainties for a given budget and/or allowable estimation errors or minimize the cost of allocation given a specific minimum requirement of observability and estimation errors. It was found that formulations that used the topological definition of observability---especially IP-based methods---lend themselves well for incorporating uncertainties related to measurement errors, topological changes, and PMU or branch outages. The numerical observability-based problem formulation---though thorough in its consideration of improving state estimation accuracy---is not always easy to formulate. In practical situations, network topological information might be lacking and/or the complexity of the optimization problem under uncertainties might be high.

Other concepts that are considered in the measurement placement for state estimation improvement are critical measurements, measurement redundancy, and infrastructure costs. The concept of degree of observability also provides a practical way to account for phased deployment in the measurement placement formulation. Other critical factors considered in the placement problem for transmission systems are:

\vspace{-.1cm}
\begin{itemize}
\item Cost of PMU: In addition to sensor cost, other costs included substation upgrade, instrument transformers such as CTs/PTs, communications costs, installation and foundation costs, network maintenance cost, data archival platforms such as phasor data concentrators, and/or selected application costs of wide-area measurement and control systems
\item Network parameters: participation factor of system states, zero-injection buses, and preexisting PMU measurements
\item PMU aspects: measurement channel limits, data availability, PMU outages, and degree of redundancy
\item Uncertainties modeled: load variations, generation or branch outage/variations, network parameters, DER outputs, transmission line outages, topology changes, and N-1/N-2 contingencies
\item Other applications in addition to state estimation: dynamic state estimation, stability assessments such as transient rotor angle stability and voltage stability, line outage or topology detection, and fault identification.
\end{itemize}

\vspace{-.2cm}
Some of these applications related to stability and security assessments typically use the concept of identifying the most critical bus (with respect to a certain stability or a security index of interest). The IP method for formulating the measurement allocation problem can easily accommodate such considerations of critical buses for sensor placement.

In works related to distribution system measurement placement, the typical strategy is to reduce the estimation errors for system states and depends on the use of pseudo-measurements and Monte Carlo simulations. This is partly because of the large node-to-measurement ratio in distribution systems and the general lack of observability---or even the lack of network metadata to model the state estimation problem using either the notion of numerical or topological observability. As a consequence, studies use a less strict definition of topological observability, and missing data are filled with pseudo-measurements or planning data. Given the uncertainties introduced because of the lack of data as well as increasing levels of distributed resources, studies also resort to probabilistic approaches to solve the estimation or stability assessment problems; hence, Monte Carlo simulations provide a simple and straightforward way to ensure that a measurement placement solution is suitable for a large percentage of time against operational uncertainties. Typical factors considered in the distribution grid measurement placement problems are:

\vspace{-.2cm}
\begin{itemize}
\item Heterogeneous measurement devices (PMUs, micro-PMUs, smart meters), multichannel measurements, and pseudo-measurements
\item Predefined number of meters and rule-based candidate selection
\item Uncertainties: load time series, DG uncertainties, measurement device uncertainty, network parameter, power flow uncertainties, topology reconfiguration, pseudo-measurement uncertainties, and critical data loss and associated costs.
\end{itemize}

\vspace{-.4cm}
\subsection{Future directions}

\begin{itemize}
\item \textbf{Consideration of leverage measurements for static state estimation in transmission grids:} The notion of leverage measurement was presented in Subsection \ref{subsec.RedundancyT}. We conjecture that the IP method and its extensions can be modified to address the issues caused by leverage measurements. One approach is to use the hat matrix $\bm{S}$ given in (\ref{eq.27d}). Recall that $\widehat{\bm{z}}=\bm{S}\bm{z}$. It follows that the diagonal element of the hat matrix, $0<S_{ii}<1$, represents the influence of the $i$-th measurement $\bm{z}$ on its estimate, $\widehat{\bm{z}}$. This fact can be used to avoid placing leverage measurements---that is, to avoid placing measurements for which $S_{ii}$ is larger than a threshold. This approach is suggested in chapter 6.3 of \cite{Abur2004}; however, $S_{ii}$ is directly related to the Mahalanobis distance, hence it is vulnerable to the masking effect \cite{Mili1985b}. Another approach is to place a cluster of measurements around locations known to create leverage measurements. This approach is suggested in \cite{Mili1991}. This remains an open research topic, and it requires further investigation.
\end{itemize}

\begin{itemize}
\item \textbf{Observability for dynamic state estimation in transmission grids:} Power system dynamic state estimation is a timely topic. Achieving (strong) observability \cite{Rouhani2017, Zongsheng2021} for dynamic state estimation represents a major effort, a line of research that has been pursued \cite{Zhang2010, Tai2013, Qi2015, Sun2011}. In all previous works, the synergy between static and dynamic state estimation, briefly discussed in \cite{Abur2016}, has been completely neglected. Also, none of the previous works considered placing \emph{merging units} in addition to phasor measurement units. An effective multistage strategy should consider that modern digital substations \cite{Hunt2019} come with merging units by standard.
\end{itemize}

\begin{itemize}
\item \textbf{Merging units:} Most work focuses on PMUs in transmission grids and micro-PMUs and smart meters in distribution grids; however, emerging high-resolution sensing and measurement systems---such as merging units that can acquire sample values at 5--10 kHz---will also play a vital role in future grids with increasing levels of dynamic and stochastic phenomena. Especially when it comes to i) developing improved dynamic models as well as substation protection and control systems and ii) detecting high-impedance faults with higher levels of inverter-based resources that typically contribute very low fault currents. The availability of such high-resolution data and their fast communications and data analytics becomes highly attractive for developing autonomous, resilient systems. Though many such dynamic phenomena are local, the question of interest for the optimal placement of merging units will focus on identifying the most important locations or substations among the millions of inverter-interconnected candidate nodes that will be key to have merging units. The availability of other high-resolution sensing and measurement systems---such as digital fault recorders, digital protective relays, and micro-PMUs---must also be considered while analyzing which location will benefit from merging units data.
\end{itemize}

\begin{itemize}
\item \textbf{Consideration of notions of stability other than voltage stability and rotor angle stability:} There is renewed interest in other notions of stability, including frequency stability and resonance stability \cite{Hatziargyriou2020}, because of the integration of converter-interfaced generation in modern power systems. These notions have never been considered in the measurement placement, and they represent an open field to be exploited. Resonance stability assessment, in particular, will require high-resolution measurements provided by merging units.
\end{itemize}

\begin{itemize}
\item \textbf{Elimination of pseudo-measurements for static state estimation in distribution grids:} The reliance on pseudo-measurements has long been the only alternative for static state estimation in distribution grids; however, the continuous increase in distributed generation is increasing the level of uncertainty associated with pseudo-measurements. The elimination of pseudo-measurements has never been considered in the measurement placement problem, and it could be an interesting direction to pursue. The progressive elimination of pseudo-measurements via a multistage formulation is of particular interest to the industry.
\end{itemize}

\begin{itemize}
\item \textbf{Open-source software:} To date, open-source software for measurement placement is not available. Power system operators are highly interested in such a computational tool, but in general they do not have in-house expertise to develop it. Moreover, it is difficult to justify the investment in a tool that will not be continuously used. This type of initiative must depart from governmental initiatives.
\end{itemize}

\begin{table}[!t]
\centering \scriptsize
\setlength{\tabcolsep}{0.5em}
\caption{Distribution grid test systems}
\begin{tabularx}{\columnwidth}{l r X}
\cmidrule[1pt]{1-3}
\rowcolor{black!10}\textbf{System name} & \textbf{No. buses} & \textbf{Published works} \\ \hline
11-node        &   11 & \cite{Ma1996} \\ \hline
IEEE 13-node   &   13 & \cite{Wang2004, Chen2016, Ali2016} \\ \hline
UKGDS \#1      &   16 & \cite{Liu2012, Liu2014} \\ \hline
Italy \#2      &   17 & \cite{Muscas2006} \\ \hline
Italy \#3      &   25 & \cite{Cataliotti2016a, Cataliotti2016b} \\ \hline
30-node        &   30 & \cite{Dag2012, Samudrala2020} \\ \hline
32-node        &   32 & \cite{Damavandi2015} \\ \hline
IEEE 33-node   &   33 & \cite{Liu2002, Chen2016, Wang2018, Teimourzadeh2019} \\ \hline
IEEE 34-node   &   34 & \cite{Baran1995, Baran1996, Leou1996, Wang2004, Brinkmann2017, Zamani2018, Jamei2018, Jamei2020} \\ \hline
IEEE 37-node   &   37 & \cite{Samudrala2020} \\ \hline
Italy \#4      &   51 & \cite{Muscas2009} \\ \hline
55-node        &   55 & \cite{Xygkis2016} \\ \hline
IEEE 69-node   &   69 & \cite{Prasad2017, Prasad2018} \\ \hline
70-node        &   70 & \cite{Damavandi2015} \\ \hline
UKGDS \#2      &   77 & \cite{Liu2012, Nusrat2012} \\ \hline
Netherlands    &   77 & \cite{Xiang2014} \\ \hline
India \#3      &   85 & \cite{Prasad2017,Prasad2018} \\ \hline
Italy \#5      &   84 & \cite{Muscas2009} \\ \hline
Italy \#6      &   95 & \cite{Pegoraro2013} \\ \hline
UKGDS \#3      &   95 & \cite{Shafiu2005, Singh2009, Singh2011, Xygkis2017} \\ \hline
119-node       &  119 & \cite{Damavandi2015, Wang2018} \\ \hline
IEEE 123-node  &  123 & \cite{Wang2004, Picallo2019, Jamei2020, Samudrala2020} \\ \hline
Brazil \#4     &  134 & \cite{Biscaro2010} \\ \hline
Brazil \#3     &  136 & \cite{Xygkis2018} \\ \hline
Brazil \#5     &  141 & \cite{Pereira2004} \\ \hline
183-node       &  183 & \cite{Samudrala2020} \\ \hline
Europe \#2     &  906 & \cite{Samudrala2020} \\ \hline 
IEEE 8500-node & 8500 & \cite{Picallo2019} \\ \hline
\cmidrule[1pt]{1-3}
\end{tabularx}
\label{tab.dist.works}
\end{table}

\begin{table}[t!]
\centering \scriptsize
\setlength{\tabcolsep}{0.5em}
\caption{Transmission grid test systems}
\begin{tabularx}{\columnwidth}{l r X}
\cmidrule[1pt]{1-3}
\rowcolor{black!10}\textbf{System name} & \textbf{No. buses} & \textbf{Published works} \\ \hline
5-bus                  &    5 & \cite{Chen2006, Donmez2011} \\ \hline
WSCC 9-bus             &    9 & \cite{Aminifar2010, Aminifar2011, Aminifar2013, Gomez2014, Qi2015, Zhang2010, Tai2013, Farach1993} \\ \hline
IEEE 14-bus            &   14 & \cite{Madtharad2003, Rakpenthai2007, Baldwin1993, Xu2004, Xu2005, Cho2001, Nuqui2002, Nuqui2005, Rakpenthai2005, deSouza2005, Chen2008b, Emami2008, Korkali2009b, Zhang2013, Korres2015, Koutsoukis2013, Kekatos2012, Caro2012, Huang2014a, Huang2014b, Khajeh2017, Kim2018, Zhang2013, Muller2016, Sodhi2015, Wang2014, Rahman2017, Yang2015, Dalali2016, Brueni1993, Denegri2002, Lien2006, Peng2006, Dua2008, Chakrabarti2008, Gou2008a, Gou2008b, Zhou2008, Abbasy2009, Chakrabarti2009, Aminifar2009, Sodhi2011, Marin2003, Aminifar2010, Hurtgen2010, Peng2010, Hajian2011, Sodhi2010, Ahmadi2011, Donmez2011, Li2011, Li2013, Jamuna2012, Zhao2012, Esmaili2013, Kavasseri2011, Gol2013, Mazhari2013, Tai2013, Gou2014, Li2014, Zhao2014, Rather2015, Singh2017, Almasabi2018, Koochi2020, Kumar2005, Madtharad2005} \\ \hline
18-bus                 &   18 & \cite{Kumar2005} \\ \hline
Longitudinal 21-bus    &   21 & \cite{Palmer1996} \\ \hline
Southern Italy         &   22 & \cite{Denegri2002} \\ \hline
IEEE 24-bus            &   24 & \cite{Muller2016, Rahman2017, Chakrabarti2008, Zhao2012, Rather2015} \\ \hline
West Bengal            &   24 & \cite{Appasani2018} \\ \hline
New Zealand            &   27 & \cite{Madtharad2005} \\ \hline
Georgia-Florida        &   29 & \cite{El-Shal1990} \\ \hline
Hydro-Qu{\'e}bec \#1   &   29 & \cite{Kamwa2002} \\ \hline
IEEE 30-bus            &   30 & \cite{Rakpenthai2007, Cho2001, Nuqui2002, Nuqui2005, Rakpenthai2005, Chen2008b, Emami2008, Emami2010, Korkali2009b, Zhang2013, Korres2015, Koutsoukis2013, Kekatos2012, Huang2014a, Khajeh2017, Kim2018, Zhang2013, Muller2016, Rahman2017, Yang2015, Mohammadi2016, Pal2017b, Dalali2016, Chakrabarti2008, Gou2008b, Zhou2008, Abbasy2009, Chakrabarti2009, Aminifar2009, Marin2003, Aminifar2010, Hajian2011, Ahmadi2011, Li2013, Jamuna2012, Shahraeini2012, Zhao2012, Mazhari2013, Wen2013, Li2014, Rather2015, Singh2017, Almasabi2018, Sun2019, Zhu2019, Koochi2020, Shi2020, Kumar2005} \\ \hline
Central Southern Italy &   38 & \cite{Denegri2002} \\ \hline
New England 39-bus     &   39 & \cite{Rakpenthai2007, Mili1990, Baldwin1993, Milosevic2003, Koutsoukis2013, Huang2014a, Khajeh2017, Muller2016, Rahman2017, Rashidi2016, Kumar2016, Denegri2002, Peng2006, Chakrabarti2008, Sodhi2011, Aminifar2010, Peng2010, Hajian2011, Sodhi2010, Jamuna2012, Mazhari2013, Tai2013, Castillo2020, Wang2020, Rather2015, Li2019, Shi2020} \\ \hline
Iran \#1               &   50 & \cite{Koochi2020} \\ \hline
IEEE 57-bus            &   57 & \cite{Rakpenthai2007, Xu2004, Xu2005, Chen2005, Chen2006, Cho2001, Nuqui2002, Nuqui2005, Korkali2009b, Zhang2013, Aminifar2013, Gomez2014, Korres2015, Koutsoukis2013, Caro2012, Khajeh2017, Kim2018, Zhang2013, Muller2016, Rahman2017, Aghaei2015, Dalali2016, Peng2006, Dua2008, Gou2008b, Zhou2008, Abbasy2009, Chakrabarti2009, Aminifar2009, Marin2003, Aminifar2010, Hurtgen2010, Hajian2011, Ahmadi2011, Aminifar2011, Li2013, Shahraeini2012, Esmaili2013, Kavasseri2011, Mazhari2013, Wen2013, Zhu2019, Shi2020} \\ \hline
Brazil \#1             &   61 & \cite{deSouza2005} \\ \hline
Hydro-Qu{\'e}bec \#2   &   67 & \cite{Kamwa2002} \\ \hline
New England 68-bus     &   68 & \cite{Sun2011, Lara-Jimenez2017, Li2019} \\ \hline
North Central U.S. and Canada & 75 & \cite{Heydt1989} \\ \hline
Simplified Italy       &   76 & \cite{Denegri2002} \\ \hline
RTS96                  &   96 & \cite{Li2013, Gomez2014, Wang2014, Almasabi2018} \\ \hline
IEEE 118-bus           &  118 & \cite{Rakpenthai2005, Mili1990, Baldwin1993, Xu2004, Xu2005, Chen2005, Chen2006, Milosevic2003, Korkali2009b, Korres2015, Koutsoukis2013, Kekatos2012, Huang2014a, Huang2014b, Khajeh2017, Kim2018, Muller2016, Rahman2017, Yang2015, Mohammadi2016, Pal2017b, Pal2014, Dalali2016, Denegri2002, Dua2008, Zhou2008, Abbasy2009, Chakrabarti2009, Aminifar2009, Marin2003, Aminifar2010, Hurtgen2010, Hajian2011, Ahmadi2011, Li2011, Azizi2012, Shahraeini2012, Esmaili2013, Kavasseri2011, Mazhari2013, Wen2013, Tai2013, Gou2014, Li2014, Wang2020, Singh2017, Almasabi2018, Sarailoo2018, Zhu2019, Koochi2020, Shi2020} \\ \hline
Italy \#1              &  129 & \cite{Denegri2002} \\ \hline
NPCC 48-machine        &  140 & \cite{Qi2015} \\ \hline
IEEE 50-machine        &  145 & \cite{Sun2011} \\ \hline
WSCC 173-bus           &  173 & \cite{Baldwin1993, Brueni1993, Denegri2002} \\ \hline
Mexico \#1             &  190 & \cite{Lara-Jimenez2017} \\ \hline
Taiwan \#1             &  199 & \cite{Mili1990, Baldwin1993} \\ \hline
Iran \#2               &  242 & \cite{Mazhari2013} \\ \hline
India \#1              &  246 & \cite{Sodhi2011, Sodhi2010, Sodhi2015, Kumar2016} \\ \hline
Taiwan \#2             &  265 & \cite{Baldwin1993} \\ \hline
USA \#1                &  270 & \cite{Nuqui2002, Nuqui2005} \\ \hline
Central America        &  283 & \cite{Pal2014} \\ \hline
298-bus                &  298 & \cite{Chakrabarti2008, Chakrabarti2009} \\ \hline
IEEE 300-bus           &  300 & \cite{Korkali2009b, Esmaili2013, Korres2015, Muller2016, Rahman2017, Yang2015, Pal2017b, Pal2014, Wen2013, Rather2015, Mohammadi2016, Zhu2019, Shi2020} \\ \hline
AEP utility company    &  360 & \cite{Nuqui2001} \\ \hline
USA \#2                &  444 & \cite{Nuqui2002, Nuqui2005} \\ \hline
Denmark                &  470 & \cite{Rather2015} \\ \hline
Iran \#3               &  529 & \cite{Azizi2012} \\ \hline
India \#2              &  996 & \cite{Pal2014} \\ \hline
Europe \#1             & 1354 & \cite{Shi2020} \\ \hline
Brazil \#2             & 1495 & \cite{Zhou2008} \\ \hline
Entergy                & 2285 & \cite{Emami2008, Emami2010} \\ \hline
Poland \#1             & 2383 & \cite{Aminifar2010, Koutsoukis2013, Huang2014a, Muller2016, Pal2017b, Dalali2016, Kavasseri2011, Shi2020} \\ \hline
Poland \#2             & 2746 & \cite{Aminifar2009, Dalali2016} \\ \hline
Poland \#3             & 3120 & \cite{Korres2015} \\ \hline
Poland \#4             & 3375 & \cite{Esmaili2013} \\ \hline
USA \#3                & 4520 & \cite{Korkali2009b} \\ \hline
Mexico \#2             & 5449 & \cite{Castillo2020} \\ \hline
USA \#4                & 8000 & \cite{Donmez2011} \\ \hline
\cmidrule[1pt]{1-3}
\end{tabularx}
\label{tab.transm.works}
\end{table}

\vspace{-.5cm}
\begin{itemize}
\item \textbf{Scalability and application to large realistic grids:} The review presented in previous sections indicates that most measurement placement work presented has been developed and tested on IEEE and other open-source test systems, with very few real system applications, e.g., the Italian network; see tables \ref{tab.dist.works} and \ref{tab.transm.works}. The important information contained in these tables is twofold: i) the performance of most existing methods has not been assessed on realistically sized systems, and ii) research groups that have the capability to work with realistically sized systems are clearly identified.

On transmission grid applications, several papers have applied their proposed measurement placement algorithms on many test systems---the IEEE 14-, 30-, 57-, and 118-bus systems are frequently used---thereby providing a means to compare the algorithms and results; however, such a practice is less common on distribution grid applications partly because of the fewer related works and level of maturity of the research. In other words, measurement placement for transmission grid applications has gotten more attention over the years than distribution grid applications, which are increasing in relevance with increasing penetrations of DERs. In general, there is a dearth of studies on large-scale, realistic systems. One possible future direction could be to apply appropriate methods to synthetic, large-scale, realistic test systems developed as part of the ARPA-E GRID DATA program \cite{Krishnan2020}. Both the open-source transmission \cite{data_transmission} and distribution \cite{data_distribution} data sets available are large enough and will be able to challenge the measurement placement algorithms. For instance, the synthetic Bay Area distribution data set models more than 4 million customers.
\end{itemize}

\vspace{-.5cm}
\begin{itemize}
\item \textbf{Planning uncertainties:} Other areas of future research can be in terms of modeling several sources of power system uncertainties. Typical considered uncertainties include PMU outage, branch outage, bad measurements, and load and DER outputs. Most of these are short-term operational uncertainties; however, given that the measurement placement problem is a planning problem, long-term uncertainties related to grid futures could also be considered, such as customer adoption levels of DERs, variations in DER sizing, displacement of conventional synchronous generators, degree of proliferation of smart inverters or control devices, and load growth (including electric vehicle penetrations). These factors can dictate the required level of observability and controllability needed for various applications, the stability drivers in the grid, and will consequently influence the locations, type, and number of measurements required.
\end{itemize}

\begin{itemize}
\item \textbf{Synergies from disparate measurement systems and scalable processing systems:} Another future direction to pursue is modeling and quantifying the synergies across multiple measurement systems. Most work reviewed in this paper considered synergies across grid electrical output measurements, e.g., PMUs, meters. Very few work has also considered weather forecasts as pseudo-measurements of variable renewable generation output. This idea can be further extended to form an optimal placement of measurements considering both the grid and weather sensing devices. Further extension could include measurements in buildings, water delivery systems, and fuel and transportation infrastructures. For instance, the optimal placement of all-sky imagers \cite{Shaffery2020} or pyranometers will be essential to forecast the net load in futures with higher penetrations of distributed solar; hence, leveraging data from smart meters, PMUs, and weather measurements will provide an economic solution for grid-edge observability. The solution strategy could also include remote sensing, such as satellite image processing. For both short-term operational states and long-term planning states, learning algorithms could be trained to estimate the grid-edge states and identify the most critical locations that will need sensing of appropriate electrical or weather parameters to ensure the system observability, estimate the states, and inform stability indices of interest. One aspect related to this direction of work is the need for centralized processing systems that can assimilate and curate all these disparate data streams and distribute them to various operations or use cases in a business system. Some utilities have already begun the development of a such a centralized data ingestion platform. The use of the Apache Spark or Kafka platforms is also gaining traction among researchers and the power industry to develop a scalable, heterogeneous data ingestion and distributed analytics platform.
\end{itemize}

\begin{itemize}
\item \textbf{Artificial intelligence, data anomaly detection and fusion with model-generated data:} In addition to fusing raw measured data, artificial intelligence techniques can add further richness to the data by using offline model-generated data, especially to capture the influence of low-probability, high-impact events and to detect unforeseen patterns. This is particularly useful for applications that are geared toward detecting anomalous data, cybersecurity intrusions, and ensuring grid resilience---something for which not many past works have designed optimal measurement placement methods, as shown from the reviews in Section \ref{section.T} and Section \ref{section.D}. Future work must look to fuse disparate measurement system data sets along with validated model-generated data through artificial intelligence. Typically information content (or entropy) is high in the region of system parameter measurements where the variability of grid reliability or resilience is high, e.g., the boundary region between acceptable and unacceptable grid performance \cite{Krishnan2011}. Given that such regions are less represented in typical grid measured conditions, analytics that will further enrich the data for such conditions through extensive offline model analysis will be needed. Such efficient sampling of training data can ensure optimal measurement placement for applications targeted toward ensuring grid resilience.

In terms of methodologies, typical methods for optimal measurement placement are optimization or heuristic based. The former is highly useful considering that planning and state estimation problems can be represented in terms of optimization; however, considering the future research directions in this area that will need to work with large-scale, realistic systems, detailed models of AC power flow equations, as well as voltage control devices---and that will need to consider both short-term and long-term uncertainties and exploit the spatial and temporal synergies across multiple types of measurements (e.g., sensing for weather, grid, building, asset health)---scalable data-driven approaches that can fuse the disparate data from myriad measurements are definitely needed. Machine learning and statistical methods, such as clustering and decision trees \cite{Krishnan2011, Bastos2020}, as well as signal processing techniques \cite{Netto2019} have already been used in the past to identify the most representative or influential network nodes to be monitored and their related measurements. Such methods are attractive especially for distribution grid applications \cite{Murphy2017}, given the lack of network models and metering infrastructure, including on the secondary side of service transformers.
\end{itemize}

\begin{itemize}
\item \textbf{Business value proposition:} For real utility or industry adoption of solutions, the value proposition of measurement placement solutions will need to consider relevance to multiple stakeholders, applications, and planning time horizons. The papers reviewed typically focus on one or two applications, and many revolve around ensuring complete or maximizing observability; however, a comprehensive placement method for a business must consider myriad high-impact applications or use cases (some of which were delineated in this paper) and also consider values from both short-term (grid operational performance) and long-term perspectives. Given that business models in utilities are evolving around the use and management of data to ensure reliable service and innovative products for increased customer satisfaction, the measurement system allocation problem must consider the value streams and applications important for a typical business model. One way to ensure this will be to work on real system data and to work closely with industry stakeholders to exploit the data streams, both live and historical archives.
\end{itemize}

\section{Conclusions}

\vspace{-.3cm}
This paper reviewed various methodologies for optimal measurement placement in transmission and distribution systems. In general, all the transmission/distribution systems have preexisting measurements, but most methods ignored this fact and placed measurements afresh, and very few methodologies considered preexisting measurements. Future methods will need to consider the synergies across the disparate sensing systems, including cutting-edge, upcoming systems, such as merging units, to improve the performance of a wide range of applications and to ensure economic plans for smart grid investments. Many methodologies---particularly the ones for distribution grids---are validated on small distribution system models of less than 150 nodes; however, the size of the practical distribution system is much larger, and the practical application of these methods is not well addressed. Most work in the open literature has adopted the IP method and its extensions for their \emph{flexibility}, \emph{scalability}, and \emph{applicability to real systems}. Among the various references cited in this paper, the works that seek the minimum-cost PMU placement for system observability and that report results in networks of 500 buses or more all rely on some variation of the IP method; see Table \ref{tab.transm.works}. Future work on the placement of measurements should consider the integration of disparate and synergistic sensing and measurement system data in addition to synthetic generated data; in information technology/operational technology system platforms, such as energy management systems and advanced distribution management systems; and/or scalable data set ingestion platforms in businesses. The use of artificial intelligence methods for sensor placement methods to counter operational and planning uncertainties as well as future business needs to ensure cybersecurity as well as resilience will gain more traction. 

\vspace{-.3cm}
\section{Acknowledgments}
{\color{white}.}

\vspace{-.5cm}
This work was authored in part by the National Renewable Energy Laboratory (NREL), operated by Alliance for Sustainable Energy, LLC, for the U.S. Department of Energy (DOE) under Contract No. DE-AC36-08GO28308. {\color{black}Funding provided by the U.S. Department of Energy Office of Electricity Advanced Grid Research and Development}. The views expressed in the article do not necessarily represent the views of the DOE or the U.S. Government. The U.S. Government and the publisher, by accepting the article for publication, acknowledge that the U.S. Government retains a nonexclusive, paid-up, irrevocable, worldwide license to publish or reproduce the published form of this work, or allow others to do so, for U.S. Government purposes. 

M. Netto acknowledges support by the NREL Director's Postdoctoral Fellowship under the Laboratory Directed Research and Development (LDRD) program at NREL.

\vspace{-.3cm}
\bibliographystyle{IEEEtran}

\end{document}